\documentclass[a4paper,11pt]{article}
\pdfoutput=1 

\usepackage{jcappub} 
\usepackage[T1]{fontenc} 

\usepackage{graphicx,amssymb,amsmath,amsthm,amsfonts,epsfig,times,bm}
\usepackage{aas_macros}
\usepackage{cleveref}

\usepackage{amsmath,amssymb}
\usepackage{tensor}

\def\be{\begin{equation}}
\def\ee{\end{equation}}

\newcommand{\bea}{\begin{eqnarray}}
\newcommand{\eea}{\end{eqnarray}}
\def\l{\left}
\def\r{\right}
\def\nn{\nonumber}

\def\f{\frac}
\def\om{\omega_{lmn}}
\def\flm{f_{lmn}}
\def\ph{\phi_{lmn}}
\def\ta{\tau_{lmn}}
\def\Qlm{Q_{lmn}}
\def\Slm{S_{lmn}}
\def\ii{{\rm i}}

\title{\boldmath Black holes and gravitational waves in models of minicharged dark matter}

\author[1,2]{Vitor Cardoso,}
\author[1]{Caio F. B. Macedo,}
\author[3,1]{Paolo Pani,}
\author[3]{Valeria Ferrari}
\affiliation[1]{CENTRA, Departamento de F\'{\i}sica, Instituto Superior T\'ecnico -- IST, Universidade de Lisboa -- UL,
Avenida Rovisco Pais 1, 1049 Lisboa, Portugal}
\affiliation[2]{Perimeter Institute for Theoretical Physics, 31 Caroline Street North
Waterloo, Ontario N2L 2Y5, Canada}
\affiliation[3]{Dipartimento di Fisica, ``Sapienza'' Universit\`a di Roma \& Sezione INFN Roma1, P.A. Moro 5, 00185, Roma, Italy}

\abstract
{
In viable models of minicharged dark matter, astrophysical black holes might be charged under a hidden $U(1)$ symmetry and are formally described by the same Kerr-Newman solution of Einstein-Maxwell theory. These objects are unique probes of minicharged dark matter and dark photons. We show that the recent gravitational-wave detection of a binary black-hole coalescence by aLIGO provides various observational bounds on the black hole's charge, regardless of its nature.
The pre-merger inspiral phase can be used to constrain the dipolar emission of (ordinary and dark) photons, whereas the detection of the quasinormal modes set an upper limit on the final black hole's charge. By using a toy model of a point charge plunging into a Reissner-Nordstrom black hole, we also show that in dynamical processes the (hidden) electromagnetic quasinormal modes of the final object are excited to considerable amplitude in the gravitational-wave spectrum only when the black hole is nearly extremal. 
The coalescence produces a burst of low-frequency dark photons which might provide a possible electromagnetic counterpart to black-hole mergers in these scenarios.
}

\begin{document}

\maketitle
\flushbottom

\tableofcontents
\section{Introduction}\label{sec:intro}

Astrophysical black holes (BHs) are considered to be electrically neutral due to quantum discharge effects~\cite{Gibbons:1975kk}, 
electron-positron pair production~\cite{1969ApJ...157..869G,1975ApJ...196...51R,Blandford:1977ds},
and charge neutralization by astrophysical plasmas. These arguments rely ---~one way or the other~--- on the huge charge-to-mass ratio of the electron\footnote{Through this work we use $G=c=\hbar=1$ units and unrationalized Gaussian units for the charge, unless otherwise stated.}, $e/m_e\approx 10^{21}$. 
Together with the celebrated BH no-hair theorems (cf. Ref.~\cite{Robinson} for a review), these arguments imply that ---~within Einstein-Maxwell theory~--- vacuum astrophysical BHs are described by a special case of the Kerr-Newman metric~\cite{KerrNewman}, namely the Kerr solution~\cite{Kerr:1963ud}. The latter is characterized only by its mass $M$ and angular momentum $J:=\chi M^2$, since the electric BH charge $Q_{\rm em}$ is assumed to be negligible in astrophysical scenarios.

On the other hand, models of minicharged dark matter (DM) predict the existence of new fermions which possess a fractional electric charge or are charged under a hidden $U(1)$ symmetry~\cite{DeRujula:1989fe,Perl:1997nd,Holdom:1985ag,Sigurdson:2004zp,Davidson:2000hf,McDermott:2010pa}. Their corresponding charge is naturally much smaller than the electron charge and their coupling to the Maxwell sector is suppressed. These minicharged particles are a viable candidate for cold DM and their properties have been constrained by several cosmological observations and direct-detection experiments~\cite{Davidson:2000hf,Dubovsky:2003yn,Gies:2006ca,Gies:2006hv,Perl:2009zz,Burrage:2009yz,Ahlers:2009kh,Haas:2014dda,Kadota:2016tqq,Gondolo:2016mrz}. 
In some other models dark fermions do not possess (fractional) electric charge but interact among each other only through the exchange of dark photons, the latter being the mediators of a long-range gauge interaction with no coupling to Standard-Model particles~\cite{Ackerman:mha}.

Although rather different, both these DM models introduce fermions with a small (electric and/or dark) charge. It is therefore natural to expect that minicharged DM can circumvent the stringent constraints that are in place for charged BHs in Einstein-Maxwell theory. In this work we will show that this is indeed the case and that even extremal Kerr-Newman BHs are astrophysically allowed in the presence of minicharged DM.

\begin{figure}[ht]
\begin{center}
\epsfig{file=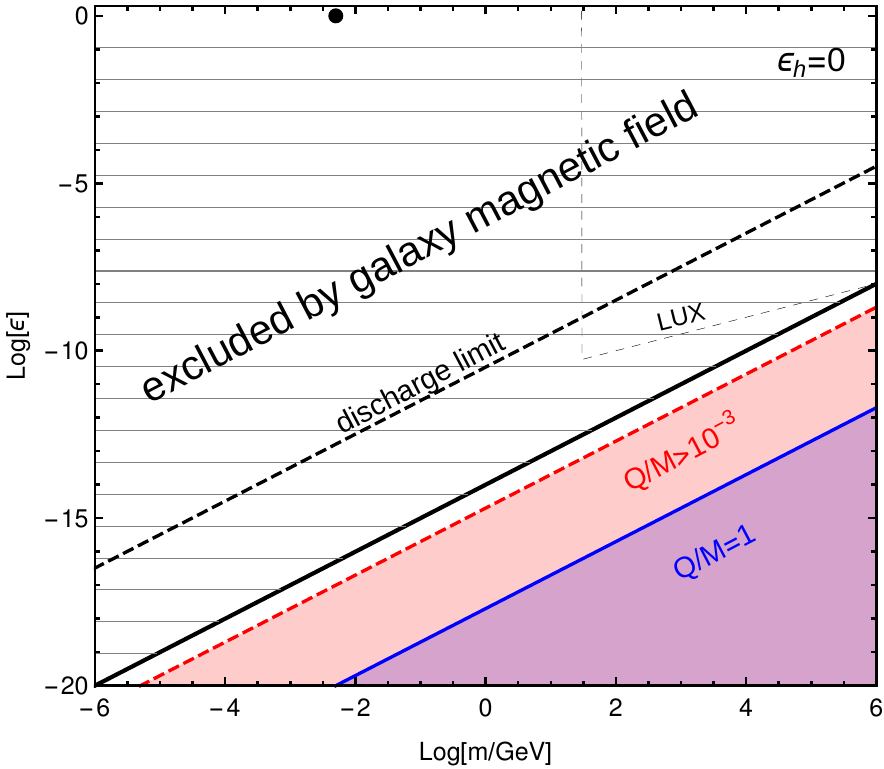,width=0.48\textwidth,angle=0,clip=true}
\epsfig{file=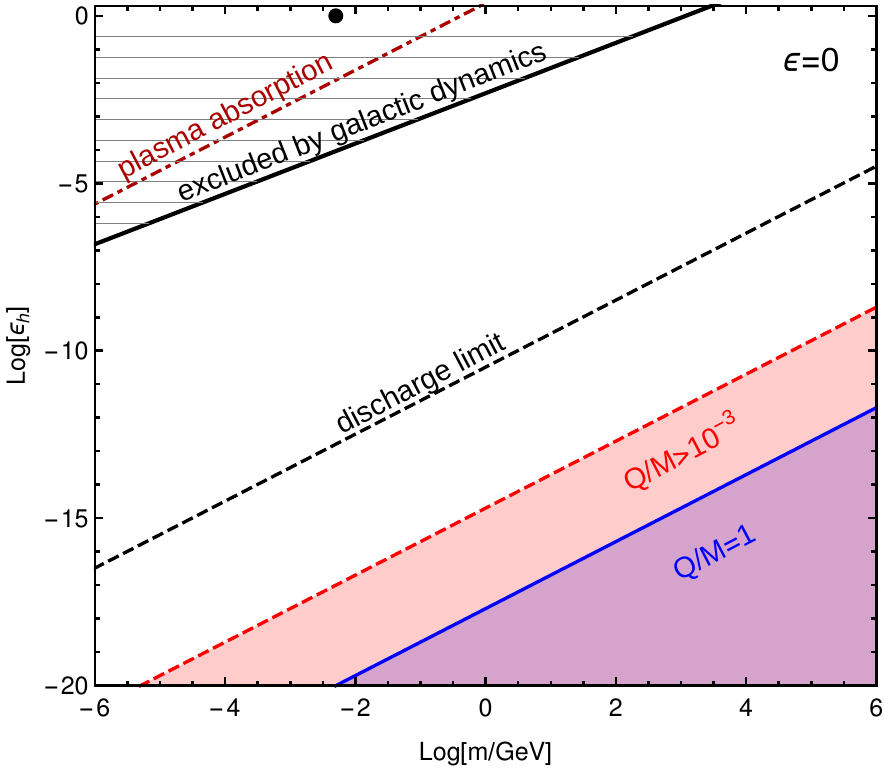,width=0.48\textwidth,angle=0,clip=true}
\caption{
The parameter space of a minicharged DM fermion with mass $m$ and charge $q=e\sqrt{\epsilon_h^2+\epsilon^2}$ (see main text for details). The left and right panels respectively show the planes $\epsilon_h=0$ and $\epsilon=0$ of the three dimensional parameter space $(m,\epsilon_h,\epsilon)$. As a reference, an electron-like particle $(m\sim 0.5\,{\rm MeV},\epsilon=1)$ is denoted by a black marker.
In each panel, the red and blue areas below the two threshold lines denote the regions where charged BHs with a charge-to-mass ratio $Q/M>10^{-3}$ and $Q/M=1$ can exist [cf. Eq.~\eqref{maxQoM}]. The region above the black dashed line is excluded because in this region extremal BHs would discharge by plasma accretion within less than the Hubble time [cf. Eq.~\eqref{taudischarge}].
{\it Left panel:} The hatched region is excluded by the effects of the magnetic fields of galaxy clusters~\cite{Kadota:2016tqq} and it is the most stringent observational constraint on the model (we also show the region excluded by the direct-detection experiment LUX~\cite{Akerib:2013tjd}, cf.~Ref.~\cite{Kadota:2016tqq} for details and other constraints).  
{\it Right panel:}  When $\epsilon=0$ our model reduces to that of DM with dark radiation~\cite{Ackerman:mha} and the region above the solid black line is excluded by soft-scattering effects on the galaxy dynamics~\cite{Ackerman:mha}. In the region above the dark red dot-dashed line hidden photons emitted during the ringdown of a $M\sim 60 M_\odot$ would be absorbed by hidden plasma of density $\rho_{\rm DM}\sim 0.4\,{\rm GeV}/{\rm cm^3}$ [cf. Eq.~\eqref{boundplasma}].
\label{fig:bounds}}
\end{center}
\end{figure}

Charged BHs are remarkably sensitive to the presence of even tiny hidden charges but are otherwise insensitive to the details of their interaction. This is a consequence of the equivalence principle of general relativity. We shall take advantage of this universality and discuss BHs charged under a fractional electric charge or under a hidden dark interaction on the same footing. Figure~\ref{fig:bounds} summarizes the main results of Section~\ref{sec:charge}, showing the parameter space of a minicharged fermion with mass $m$ and charge $q=e\sqrt{\epsilon_h^2+\epsilon^2}$ in which astrophysical charged BHs can exist (here and in the following $\epsilon_h$ and $\epsilon$ are the fractional hidden and electric charges of the dark fermion, respectively). Interestingly, such region does not overlap with the region excluded by direct-detection experiments and by cosmological observations.

Having established that charged astrophysical BHs can exist in theories of minicharged DM, we proceed to study their gravitational-wave (GW) signatures in Section~\ref{sec:GW}. We consider the coalescence of a binary BH system similar to GW150914, the event recently detected by the LIGO/Virgo Collaboration using the GW interferometers aLIGO~\cite{Abbott:2016blz}, and show that different phases of the coalescence can be used to constrain the $U(1)$ charge of the BHs in the binary and of the final BH produced in the post-merger phase. Finally, we explore the excitation of the ringdown modes and the total radiated energy during the collision of unequal-mass BHs, in a perturbative approach, and show that they are in good agreement with previous, restricted results in Numerical Relativity, as well as with simplified flat-space calculations.

\section{Charged BHs in minicharged DM models\label{sec:charge}}

\subsection{Setup}
We consider the following classical Lagrangian~\cite{Holdom:1985ag} 
\begin{equation}
 {\cal L} = \sqrt{-g}\left(\frac{R}{16\pi}- \frac{1}{4}F_{\mu\nu} F^{\mu\nu}-\frac{1}{4}B_{\mu\nu} B^{\mu\nu}+ 4\pi e j_{\rm em}^\mu A_\mu+ 4\pi  e_h j_h^\mu B_\mu +4\pi  \epsilon e j_h^\mu A_\mu\right)\,,\label{Lagrangian}
\end{equation}
where $F_{\mu\nu}:=\partial_\mu A_\nu-\partial_\nu A_\mu$ and $B_{\mu\nu}:=\partial_\mu B_\nu-\partial_\nu B_\mu$ are the field strengths of the ordinary photon and of the dark photon, respectively, $j_{\rm em}^\mu$ and $j_h^\mu$ are the electromagnetic (EM) and the hidden number currents, $e$ is the electron charge, and $e_h$ is the gauge coupling of the hidden sector. The model~\eqref{Lagrangian} describes a theory in which a charged fermion is coupled to ordinary photons with coupling $\epsilon^2 e^2$ and to dark photons with coupling $e_h^2:=\epsilon_h^2 e^2$. The parameters $\epsilon$ and $\epsilon_h$ are free\footnote{In the model studied in Refs.~\cite{Holdom:1985ag,Burrage:2009yz}, $\epsilon/\epsilon_h=\tan\theta$ defines the kinetic mixing angle $\theta$ of the photon fields before the diagonalization leading to the Lagrangian~\eqref{Lagrangian} in which $F_{\mu\nu}$ and $B_{\mu\nu}$ are decoupled (cf. Refs.~\cite{Holdom:1985ag,Burrage:2009yz} for details). In this case, the effective electron charge is $e\cos\theta$. This kinetic coupling is constrainted by arguments related to Big Bang nucleosynthesis, cosmic microwave background, and large-scale structure formation (cf., e.g., Ref.~\cite{Foot:2014uba}). For simplicity we neglect such coupling here and consider the Lagrangian~\eqref{Lagrangian} as fundamental. This corresponds to the model studied in Refs.~\cite{Holdom:1985ag,Burrage:2009yz} when $\theta\to0$.}.

The field equations arising from the Lagrangian~\eqref{Lagrangian} read 
\begin{eqnarray}
 \nabla_\mu F^{\mu\nu}&=&-4\pi e (j^\nu_{\rm em}+\epsilon j_h^\nu)\,, \label{maxwell}\\
 \nabla_\mu B^{\mu\nu}&=&- 4\pi e_h  j_h^\nu\,, \label{darkmaxwell} \\
 G_{\mu\nu}&=&8\pi(T_{\mu\nu}^{\rm em}+T_{\mu\nu}^h)\,, \label{einstein}
\end{eqnarray}
where we defined the effective stress-energy tensors
\begin{eqnarray}
 T_{\mu\nu}^{\rm em} &:=& F_{\mu\alpha}F^\alpha_\nu-\frac{1}{4}g_{\mu\nu}F^2-8\pi e A_{(\mu}  [j_{\nu)}^{\rm em}+\epsilon j_{\nu)}^h]\,, \\
 T_{\mu\nu}^h &:=& B_{\mu\alpha}B^\alpha_\nu-\frac{1}{4}g_{\mu\nu}B^2 -8\pi e_h B_{(\mu}  j_{\nu)}^h\,,
\end{eqnarray}
for the standard Maxwell field and for the dark photon, respectively. The continuity equations $\nabla_\nu j^\nu_{\rm em}=0=\nabla_\nu j^\nu_h$ for the standard and hidden currents follow directly from Eqs.~\eqref{maxwell} and \eqref{darkmaxwell}. Note that a hidden electron carries both electric charge $\epsilon e$ and hidden charge $e_h$.  

In the absence of currents ($j^\mu_{\rm em}=j^\mu_h=0$), the most general stationary solution~\cite{Robinson} of the above field equations is the Kerr-Newman~\cite{KerrNewman} metric with total charge $Q=\sqrt{Q_{\rm em}^2+Q_h^2}$, where the (EM and hidden) BH charges are defined from the solution of the (standard and hidden) Maxwell equations in static equilibrium, 
\begin{eqnarray}
 B_{tr}&=&\frac{Q_h}{r^2} \,,\\
 F_{tr}&=&\frac{Q'}{r^2}+\epsilon\frac{e}{e_h} B_{tr} := \frac{Q_{\rm em}}{r^2}
\end{eqnarray}
where $Q_{\rm em}=Q'+\epsilon\frac{e}{e_h} Q_h$. When the BH is spinning the presence of a charge induces a magnetic field along the angular directions~\cite{KerrNewman}.  Note that the hidden current $j^\nu_h$ provides both electric charge and hidden charge to the BH.
Thus, a standard electron and a dark electron feel respectively the force
\begin{eqnarray}
 F_{\rm em}&:=& e F_{tr} = e \frac{Q_{\rm em}}{r^2} \,,\\
 F_h &:=& e_h B_{tr}+\epsilon e F_{tr}= \frac{e_h Q_h+ \epsilon e Q_{\rm em}}{r^2}\,.
\end{eqnarray}
If the BH acquires charge only through the accretion of a hidden current (i.e., if $j^{\nu}_{\rm em}=0$, a condition that is enforced by several considerations as discussed in Sec.~\ref{sec:theobounds}), then $Q'=0$ and the EM and hidden charge are proportional to each other, $Q_{\rm em}=\frac{\epsilon}{\epsilon_h} Q_h$. In this case the BH charge reads $Q=Q_h \sqrt{1+\epsilon^2/\epsilon_h^2}$, and the force felt by a hidden electron reduces to
\begin{equation}
 F_h=e_h \frac{Q_h}{r^2}\left(1+\frac{\epsilon^2 }{\epsilon_h^2 }\right)= \frac{q Q}{r^2}\,, \label{forceDM}
\end{equation}
where we have defined the effective charge of a dark fermion in this model, $q:=e\sqrt{\epsilon_h^2+\epsilon^2}$.

Observational bounds on minicharged DM typically constrain the coupling to Standard-Model particles, especially the coupling $\epsilon$ to ordinary photons. As such, they are insensitive to the coupling $e_h$ which is indeed typically neglected\footnote{When $\epsilon_h\to0$ the model~\eqref{Lagrangian} simply describes a dark fermion coupled to ordinary photons with coupling $\epsilon^2 e^2\ll e^2$.}. On the other hand, the dark coupling $e_h$ plays a crucial role in models of dark radiation~\cite{Ackerman:mha}. Gravitational tests do not require ordinary photons as mediators and are indeed sensitive to the entire parameter space $(\epsilon_h,\epsilon)$. In particular, the effects we are going to discuss are present also when $\epsilon=0$, i.e. when DM does not couple to Standard-Model particles, as in dark radiation models~\cite{Ackerman:mha}.

\subsection{Theoretical bounds on the charge-to-mass ratio of astrophysical BHs} \label{sec:theobounds}
There are several mechanisms that conspire to limit the electric charge of astrophysical BHs.
One is purely kinematical. Take a BH with mass $M$ and electric charge\footnote{We are now considering the standard scenario in which the electric charge is produced by an ordinary current $j_{\rm em}^\mu$. In other words, $Q_h=0$ and $Q_{\rm em}=Q'$ as defined in the previous section.} $Q_{\rm em}$ and throw in a low-energy electron of charge $e$ and mass $m_e$. For the electron to be absorbed by the BH, then the (classical) electric force must satisfy
\be
e Q_{\rm em}\leq M m_e\,,
\ee
which can be written in terms of the dimensionless charge-to-mass ratio of the BH as
\be
\frac{Q_{\rm em}}{M}\leq \frac{m_e}{e}\approx5\times 10^{-22}\,. \label{boundforce}
\ee
%
These numbers change if the particle is thrown at large velocities, but show that the maximum charge-to-mass ratio $Q_{\rm em}/M$ is typically very small.
In addition, BHs can be neutralized by surrounding plasma. If the eletrical force overwhelms the gravitational force, charge separation can occur and the BH charge can be neutralized by particles of opposite charge. For an extremal BH with $Q_{\rm em}= M$, the total number $N$ of elementary charges (each with charge $q$ and mass $m$) that it needs to accrete from the surrounding plasma to be neutralized is~\cite{1975ARA&A..13..381E}
\be
N\sim 10^{39}\left(\frac{e}{q}\right)\left(\frac{M}{M_{\odot}}\right)\,.
\ee
This corresponds to a plasma mass of 
\be
\frac{M_{\rm plasma}}{M}\sim10^{-18}\,\left(\frac{m}{m_p}\right)\left(\frac{e}{q}\right)\,, \label{Mplasma}
\ee
with $m_p$ being the proton mass. This plasma mass is easily available under the form of interstellar matter within a small region surrounding the BH~\cite{Eardley:1975kp}. To estimate the time needed to accrete an amount of plasma such that an extremal BH is discharged, let us assume that plasma accretion occurs at the same rate of gas accretion from an ordinary accretion disk. 
In the most conservative scenario mass accretion occurs at the Eddington rate, $\dot M_{\rm Edd}=2.2\times 10^{-8} (M/M_\odot) M_\odot\,{\rm yr}^{-1}$, corresponding to a discharge time scale
\begin{equation}
 \tau_{\rm discharge}\sim 5\times 10^{-11}\left(\frac{m}{m_p}\right)\left(\frac{e}{q}\right) \,{\rm yr}\,. \label{taudischarge0}
\end{equation}
For $q=e$ and $m=m_e$, $\tau_{\rm discharge}\sim 8\times 10^{-7}\,{\rm s}$.  The above results show that ---~in Einstein-Maxwell theory~--- it is difficult to charge a BH past $Q_{\rm em}/M\sim 10^{-21}$ (in geometric units), and that -- if such BH ever acquires a charge -- it discharges very quickly.

Another possibility to form charged BHs is if the latter were born through the gravitational collapse of (charged) stars.
Self-gravitating stars are globally charged due to pressure effects, as shown by Eddington, Rosseland and others~\cite{Eddington,1924MNRAS..84..720R,1978ApJ...220..743B}: in a nutshell,
lighter charges (electrons) are easily kicked out of the star by pressure effects whereas heavy ions (protons) are stuck in the interior. The calculation, which proceeds by assuming thermal equilibrium
for positive and negative charges, yields the following result for the charge of a star~\cite{Eddington,1924MNRAS..84..720R,1978ApJ...220..743B},
%
\be
\frac{Q_{\rm star}}{M}\sim \frac{m_2-m_1}{q_2-q_1}\,,
\ee
where the stellar material is assumed to be composed mainly of two species of ions with charge and mass $(m_1,q_1)$ and $(m_2,q_2)$, respectively. For standard Maxwell theory, the charge-to-mass ratio in the star is of the order of Eq.~\eqref{boundforce}.
Thus, stars are typically charged with $\approx100\,{\rm C}$. It is reasonable to expect that, if they collapse to a BH, this small charge will remain hidden behind the horizon, giving rise to a (very weakly) charged BH with $Q_{\rm em}/M\approx 10^{-19}(M_\odot/M)$.

The above discussion imposes an extremely stringent upper bound on the EM charge of astrophysical BHs. To fulfill this bound, we assume that astrophysical BHs do not accrete ordinary charges and, therefore, we set the current $j_{\rm em}^\nu$ to zero. Thus, in the rest of this work we can safely assume that the force felt by a putative hidden electron is given by Eq.~\eqref{forceDM}.

The above discussion also shows that the bounds on the charge-to-mass ratio of astrophysical BHs become much less stringent in minicharged DM models. For a hidden electron with effective charge $q$ and mass $m$, Eq.~\eqref{boundforce} becomes
\begin{equation}
\frac{Q}{M}\leq \frac{m}{q}\,, \label{boundforceDM}
\end{equation}
which can be less stringent than unity\footnote{We recall that the charge of a nonspinning charged BH is bounded by $Q/M\leq1$ in our units, where the inequality is saturated in the extremal case.} depending on the parameters $m$ and $q$. Likewise, a BH with hidden charge surrounded by a plasma of hidden electrons can be neutralized by accreting a mass $M_{\rm plasma}$ given by Eq.~\eqref{Mplasma}. If $q\ll e$ or $m\gg m_p$, $M_{\rm plasma}$ might be a considerable fraction of the BH mass so that a charged BH would be difficult to discharge in this scenario. Furthermore, DM being almost collisionless it does not form accretion disks around compact objects so that dark plasma is accreted at the Bondi rate for collisionless fluids,
\begin{equation}
 \dot M_{\rm Bondi}\sim 4\pi \rho \frac{M^2}{v^3}\,, 
\end{equation}
where $\rho$ is the local DM density and $v$ is the relative velocity between DM and the BH. Using the above equation and Eq.~\eqref{Mplasma}, we estimate the discharge time
\begin{equation}
 \tau_{\rm discharge} \sim 0.4\left(\frac{v}{220\,{\rm km/s}}\right)^3\left(\frac{0.4\,{\rm GeV/cm^3}}{\rho}\right)\left(\frac{10 M_\odot}{M}\right)\left(\frac{m}{m_p}\right)\left(\frac{e}{q}\right) \,{\rm yr}\,, \label{taudischarge}
\end{equation}
which is much longer than Eq.~\eqref{taudischarge0} since the DM density is low. In the above expression we have normalized $\rho$ and $v$ by their typical local values.
Thus, the discharge process is much slower than for ordinary plasma. The black dashed line in Fig.~\ref{fig:bounds} shows the threshold $\tau_{\rm discharge}=1.4\times 10^{10}\,{\rm yr}$ obtained from Eq.~\eqref{taudischarge} in the $(m,q)$ parameter space. In the region below this line, the time scale to discharge an extremal BH through accretion would be much longer than the Hubble time, which we assume as a very conservative limit.

Furthermore, BHs with a large charge also have a large electric field close to the horizon. Such electric field is prone to produce spontaneous pair production via the Schwinger mechanism~\cite{Schwinger:1951nm}. This effect becomes relevant when the work done by the electric field on a Compton wavelength is of the order of the rest mass of the lightest particle~\cite{Gibbons:1975kk},
\be
\frac{qE}{m}\sim m \,.
\ee
For electric fields of order $E\sim Q/M^2$, Schwinger discharge is important unless 
\be
\frac{Q}{M}\leq \frac{Mm^2}{q}\sim 10^{-5}\frac{M}{M_{\odot}}\left(\frac{m}{m_e}\right)^2\frac{e}{q}\,. \label{boundpairDM}
\ee
It is clear that the upper bound~\eqref{boundpairDM} is much less stringent than the bound~\eqref{boundforceDM} whenever $M m\gg1$, i.e. for any $m\gg 10^{-10}(M_\odot/M)\,{\rm eV}$. In the rest of this work we shall focus on this regime, i.e. we do not consider ultralight DM whose Compton wavelength is larger than the gravitational radius of the BH.

To summarize, in minicharged DM models Eq.~\eqref{boundforceDM} provides an intrinsic constraint on the maximum charge-to-mass ratio $Q/M$ of astrophysical BHs. This constraint can be also turned around to determine the region in the $(m,q)$ plane in which BHs with a certain $Q/M$ can exist, namely
\begin{equation}
 \frac{q}{e} < 2\times 10^{-18}\left(\frac{m}{{\rm GeV}}\right)\left(\frac{M}{Q}\right)\,.\label{maxQoM}
\end{equation}
This region is shown in Fig.~\ref{fig:bounds} for two different values of $Q/M$. Interestingly, existing constraints on minicharged DM models do not rule out charged BHs. On the contrary, even extremal BHs with charge-to-mass ratio $Q/M\sim 1$ are allowed, even when $e_h=0$. We stress that EM-based constraints on minicharged DM models are insensitive to the coupling $e_h$ and that charged BHs can exist also when $\epsilon=0$, a regime which cannot be ruled out by EM observations. Finally, the colored regions shown in Fig.~\ref{fig:bounds} lay well below the black dashed threshold line for plasma discharge [cf. Eq.~\eqref{taudischarge}], so that charged BHs should not easily discharge in this scenario.

\section{Gravitational-wave tests of charged BHs in minicharged DM models\label{sec:GW}}
%

The recent GW detection of a binary BH coalescence by aLIGO~\cite{Abbott:2016blz} has given us access to the strong-field/highly-dynamical regime of the gravitational interaction. In this regime, precision GW measurements can be used to develop BH-based tests of fundamental physics~\cite{CQGFocus,Berti:2015itd}. It is therefore natural to investigate whether present and upcoming GW observations can be used to constrain a putative hidden charge of astrophysical BHs in minicharged DM scenarios.

The GW-driven coalescence of a compact binary can be characterized by three phases~\cite{Buonanno:2006ui,Berti:2007fi,Sperhake:2011xk}: the inspiral, the merger and the ringdown. The inspiral corresponds to large orbital separations and is well approximated by post-Newtonian theory; the merger phase corresponds to the highly nonlinear stage right before and after merger, and can only be described accurately through numerical simulations of the full Einstein's equations; finally, the ringdown phase corresponds to the relaxation of the highly-deformed end-product to a stationary, equilibrium solution of the field equations, and can be described by BH perturbation theory~\cite{Sperhake:2011xk,Berti:2009kk,Blanchet:2013haa}. In the rest of this paper we will focus on each of these phases, with various degrees of approximation, in order to estimate the effect of a hidden BH charge in the GW signal.

\subsection{The inspiral of two charged BHs in minicharged DM models}\label{sec:inspiral}

We model the initial inspiral of the BH binary by considering two point charges, $q_1=e\sqrt{\epsilon_{h,1}^2+\epsilon_1^2}$ and $q_2=e\sqrt{\epsilon_{h,2}^2+\epsilon_2^2}$, with masses $m_1\leq q_1 /\sqrt{G}$ and $m_2\leq q_1 /\sqrt{G}$, respectively\footnote{For clarity in this section we reinsert factors of $G$ and $c$.}. To leading order, by using Eq.~\eqref{forceDM}, the motion is governed by the equation
\begin{eqnarray}
 m_i \ddot{\mathbf{r}}_i &=& \pm \frac{G m_1 m_2}{r^3}\mathbf{r}\mp\frac{q_1 q_2}{r^3}\mathbf{r}\,,
\end{eqnarray}
where the upper (lower) sign refers to $i=1$ ($i=2$), $\mathbf{r}_i$ is the position vector of the mass $m_i$, $\mathbf{r}:= \mathbf{r}_2-\mathbf{r}_1$ is the relative position of the bodies. If we define $q_i:=\lambda_i m_i \sqrt{G}$, from the equation above it is clear that the problem can be mapped into a standard Keplerian motion of two uncharged particles where $G\to G_{\rm eff}:=G(1- \lambda_1\lambda_2)$. 

Let us assume circular orbits for simplicity\footnote{Orbits are circularized under GW-radiation reaction, so it is natural to expect that the last stages in the life of a compact binary are quasi-circular~\cite{Peters:1964zz}.}. An (electric or hidden) charge in circular motion emits dipolar radiation governed by Larmor's formula. The dipolar energy flux dominates over the quadrupolar GW flux at large distances, so it might play an important role in the early inspiral.
Because both particles are charged under the standard Maxwell field and under the hidden field, the dipolar flux 
consists of two copies of Larmor's energy dissipation~\cite{Landau} with different couplings, in addition to the 
standard GW energy flux~\cite{PoissonWill}. Namely, 
\begin{eqnarray}
\frac{dE_{\rm em}}{dt}&\sim& \frac{2}{3} \left(\frac{\epsilon_1e}{m_1} -\frac{\epsilon_2e}{m_2}\right)^2 \frac{G_{\rm eff}^2 m_1^2 m_2^2}{c^3R^4}\,, \label{EdotEM}\\
\frac{dE_h}{dt}&\sim& \frac{2}{3} \left(\frac{\epsilon_{h,1}e}{m_1} -\frac{\epsilon_{h,2}e}{m_2}\right)^2 \frac{G_{\rm eff}^2 m_1^2 m_2^2}{c^3R^4}\,, \label{EdotH}\\
\frac{dE_{\rm GW}}{dt}&\sim& \frac{32}{5c^5}\eta^2  \frac{GG_{\rm eff}^3 M^5}{ R^5}\,,\label{EdotGW}
\end{eqnarray}
where $R$ is the radius of the orbit, $M=m_1+m_2$ and $\eta=m_1m_2/M^2$. Note that if the charge-to-mass ratio of the 
two objects is the same, the corresponding dipole term is zero and (only in this case) the leading term would be 
quadrupolar like the GW flux. It is convenient to write the total dipolar flux, $\frac{dE_{\rm dip}}{dt}=\frac{dE_{\rm 
em}}{dt}+\frac{dE_h}{dt}$, as 
\begin{equation}
 \frac{dE_{\rm dip}}{dt}:= \frac{2}{3} \zeta^2 \frac{G_{\rm eff}^3 m_1^2 m_2^2}{c^3 R^4}\,, \label{Edotdip}
\end{equation}
where
\begin{equation}
 \zeta^2:=\left(\frac{\epsilon_{h,1}e}{m_1\sqrt{G_{\rm eff}}} -\frac{\epsilon_{h,2}e}{m_2\sqrt{G_{\rm eff}}}\right)^2 +\left(\frac{\epsilon_1e}{m_1\sqrt{G_{\rm eff}}} -\frac{\epsilon_2e}{m_2\sqrt{G_{\rm eff}}}\right)^2 \,. \label{zeta}
\end{equation}
From the above equation it is clear that the dipolar emission in the inspiral phase depends only on the combination $\zeta$.
Note that the latter is nonvanishing even when $\epsilon_i=0$, provided $e_{h,1}$ or $e_{h,2}$ are nonzero. On the other hand, $\zeta\sim0$ if the two BHs have similar charge-to-mass ratios.

The inspiral phase of GW150914 was compatible with the prediction of post-Newtonian theory~\cite{Abbott:2016blz,TheLIGOScientific:2016src}. This implies that any putative dipolar contribution to the energy flux must be small.
In the limit $\zeta\ll1$, the condition $\frac{dE_{\rm dip}}{dt}\ll \frac{dE_{\rm GW}}{dt}$ can be written as
\begin{equation}
 R\ll 480 \frac{G M}{c^2} \left(\frac{0.1}{\zeta}\right)^2\,, \quad \Omega\gg 0.3 {\rm Hz} 
\sqrt{\frac{G_{\rm eff}}{G}}\left(\frac{60 M_\odot}{M}\right)\left(\frac{\zeta}{0.1}\right)^3\,,
\end{equation}
where $\Omega$ is the orbital frequency and we assumed $m_1\sim m_2$. This simple estimates suggest that the dipolar correction might be small when the binary enters the aLIGO band.

To quantify the effect of the dipolar energy loss more precisely, we can compute the GW phase associated with such effect through a simple quasi-Newtonian evolution~\cite{MaggioreBook}.
The binding energy of the two-body system is $E=-G_{\rm eff} m_1 m_2/R$. By assuming an adiabatic approximation, $\dot E 
= -dE_{\rm GW}/dt-dE_{\rm dip}/dt$, we can obtain a differential equation for the orbital radius $R=R(t)$. This equation 
can be solved analytically in the limit $|dE_{\rm dip}/dt|\ll |dE_{\rm GW}/dt|$, i.e. when the dipolar loss is a small 
correction compared to the quadrupolar GW flux. In this case, a standard procedure~\cite{MaggioreBook} allows us to 
compute the amplitude and the phase of the quadrupolar GWs emitted by the system through an adiabatic evolution. The 
phase of the ``$+$'' GW polarization reads 
\begin{eqnarray}
 \Psi_+(f) &=& 2\pi f t_c-\Phi_c+\frac{3}{128}{\frac{G_{\rm eff}}{G}}\left(\frac{G_{\rm eff} {\cal M}}{c^3}\pi 
f\right)^{-5/3}\left[1-\frac{5 }{84} \eta^{2/5}\zeta^2{\frac{G_{\rm eff}}{G}}\left(\frac{G_{\rm eff} {\cal M}}{c^3} \pi 
f\right)^{-2/3} \right. \nn\\
 &&\left.+\left(\frac{3715}{756}+\frac{55}{9}\eta\right)\eta^{-2/5}\left(\frac{G_{\rm eff} {\cal M}}{c^3}\pi f\right)^{2/3} \right]\,, \label{phaseGW}
\end{eqnarray}
where $t_c$ and $\Phi_c$ are the time and phase at coalescence, $f$ is the GW frequency and ${\cal M}:=M\eta^{3/5}$ is 
the chirp mass.
When $\epsilon_i=e_{h,i}=0$, then $G_{\rm eff}=G$, $\zeta=0$, and Eq.~\eqref{phaseGW} yields the standard leading-order result, to which we added the first next-to-leading order post-Newtonian term in the second line of the above equation.

\begin{figure}[ht]
\begin{center}
\epsfig{file=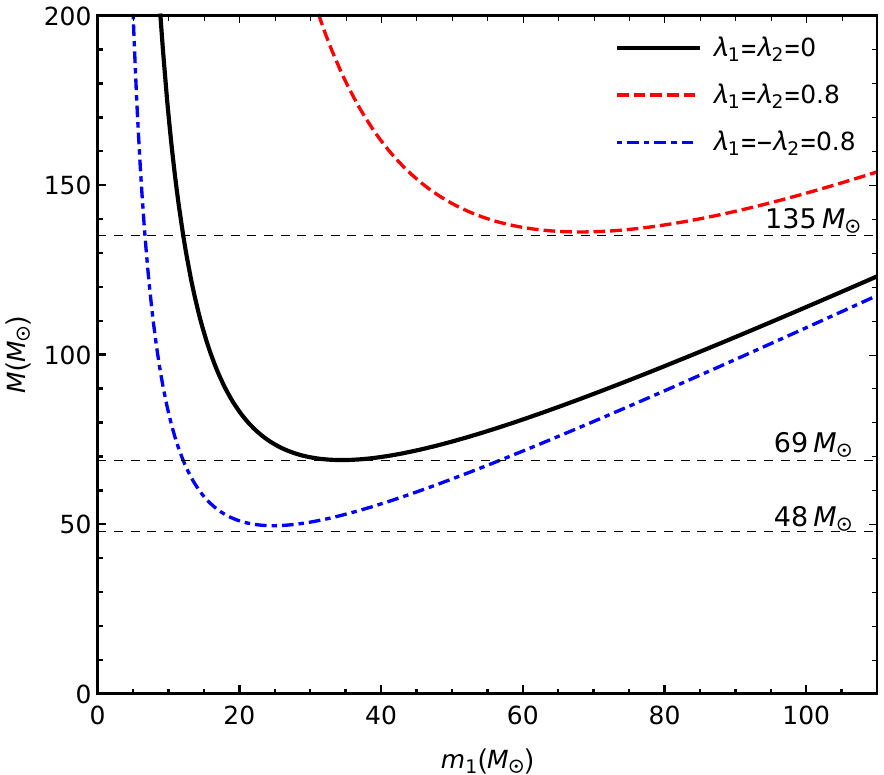,width=7.8cm,angle=0,clip=true}
\caption{Total mass $M=m_1+m_2$ of a binary system as a function of $m_1$ obtained from Eq.~\eqref{Mchirp} for fixed ${\cal M}_{\rm measured}\approx 30 M_\odot$ and for different values of $\lambda_i$. The black curve corresponds to the uncharged case~\cite{Abbott:2016blz}. Note that when $\lambda_1\lambda_2<0$ the total mass of the system can be significantly smaller than in the uncharged case.
\label{fig:Mchirp}}
\end{center}
\end{figure}

However, when at least one of the parameters $\epsilon_i$, $e_{h,i}$ are nonzero, we obtain two types of corrections. 
The first one is a rescaling of Newton's constant, which affects also the Newtonian result. This correction is present 
even if $\zeta=0$, provided $\lambda_i\neq0$.
Because to leading order the GW phase depends on the combination $(G_{\rm eff}/G)^{-2/3}(G{\cal M})^{-5/3}$, 
a rescaling of Newton's constant is degenerate with the measurement of the chirp mass. Extracting the latter from the 
Newtonian GW phase obtained by neglecting charge effects would yield a result that is rescaled by a factor $(G_{\rm 
eff}/G)^{2/3}$ relative to the real chirp mass of the system, namely
\begin{equation}
 {\cal M} := M\eta^{3/5} = \frac{{\cal M}_{\rm measured} }{(1-\lambda_1\lambda_2)^{2/3}}\,, \label{Mchirp}
\end{equation}
Figure~\ref{fig:Mchirp} shows the total mass $M$ as a function of $m_1$ for fixed ${\cal M}_{\rm measured}\approx 30 M_\odot$~\cite{Abbott:2016blz} and for different values of $\lambda_i$. When $\lambda_i=0$ we recover the standard result, namely a minimum total mass $M_{\rm min}\approx 69 M_\odot$. However, significant changes occur if $0.1\lesssim |\lambda_i| <1$. In particular, when $\lambda_1\lambda_2<0$ the effective Newton's constant is larger and the real total mass of the system can be significantly smaller than in the uncharged case. This property is intriguing since it shows that neglecting charge effects might systematically lead to overestimate the measured BH masses.

\begin{figure}[ht]
\begin{center}
\epsfig{file=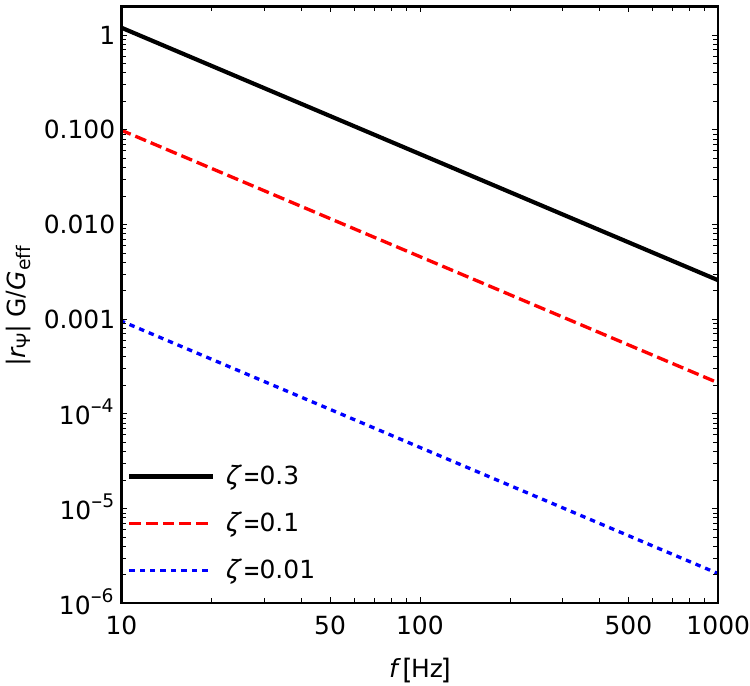,width=7.8cm,angle=0,clip=true}
\caption{Ratio $r_\Psi$ (normalized by $G_{\rm eff}/G$) between the hidden-charge-induced GW phase and the first 
post-Netwonian correction [cf. Eq.~\eqref{R}] for the quasicircular inspiral of two charged masses with $m_1=m_2=30 
M_\odot$ and different values of the coupling $\zeta$ [cf. Eq.~\eqref{zeta}] as a function of the GW frequency. Note 
that $\zeta\approx0.3$ saturates the bound presented in Eq.~\eqref{boundzetaLIGO}.
\label{fig:phase}}
\end{center}
\end{figure}

The other correction appearing in Eq.~\eqref{phaseGW} is the second term inside the square brackets in Eq.~\eqref{phaseGW}. The latter is larger at small frequencies, as expected, and in fact resembles the leading-order correction for neutron-star binaries in scalar-tensor theories\footnote{For scalar tensor theories the dipole contribution is proportional to the sensitivities, which vanish identically for BHs~\cite{Berti:2015itd}.}.
The ratio between the second term and the third term inside the square brackets in Eq.~\eqref{phaseGW}, namely 
\begin{equation}
 r_\Psi:= -\frac{18\eta^{4/5}}{743+924\eta}\zeta^2\frac{G_{\rm eff}}{G}\left(\frac{G_{\rm eff} {\cal M}}{c^3} \pi 
f\right)^{-4/3}\,,\label{R}
\end{equation}
is shown in Fig.~\ref{fig:phase} as a function of the GW frequency for a typical inspiral. For $f>30\,{\rm Hz}$, the charge-induced corrections are at least $\sim 0.03$ times smaller than the first post-Newtonian term when $\zeta\lesssim0.1$, but they can be as large as $25\%$ when $\zeta\approx 0.3$.

%


Another relevant quantity is the number of cycles spent in the detector bandwidth, ${\cal N}:=\int_{f_{\rm min}}^{f_{\rm 
max}} df \frac{f}{\dot f}$. In our case we obtain 
\begin{eqnarray}
 {\cal N} &=& \frac{1}{32\pi^{8/3}}\frac{G_{\rm eff}}{G}\left(\frac{G_{\rm eff}{\cal M}}{c^3}\right)^{-5/3}\left(f_{\rm 
min}^{-5/3}-f_{\rm max}^{-5/3}\right)\left[1-\delta_{\cal N}\right] \,, \nn
\end{eqnarray}
where
 \begin{eqnarray}
 \delta_{\cal N}=\frac{25}{336}\eta^{2/5}\zeta^2 \frac{G_{\rm eff}}{G}\left(1+\frac{f_{\rm min}^{5/3}}{f_{\rm 
max}^{5/3}}\right) \left(\frac{G_{\rm eff} {\cal M} }{c^3} \pi f_{\rm min} \right)^{-2/3} \,, \nn \\ 
\end{eqnarray}
and we have also expanded for $f_{\rm max}\gg f_{\rm min}$ to simplify the final expression. In the small-charge limit, for $f_{\rm max}\sim 100\,{\rm Hz}$ and $f_{\rm min}\sim 30\,{\rm Hz}$, we obtain
\be
\delta_{\cal N} \approx 0.01 \left(\frac{\zeta}{0.1}\right)^2\,.
\ee
Therefore, dipolar effects change the number of cycles relative to the Newtonian case by a few percent when $\zeta\approx 0.1$ and by less than $0.01\%$ when $\zeta<0.01$. On the other hand, these corrections become important at smaller frequencies and might produce detectable effects for space-based interferometers such as eLISA~\cite{AmaroSeoane:2012je}.

Very recently, Ref.~\cite{Barausse:2016eii} performed a detailed analysis to derive GW-based constraints on generic 
dipolar emissions in compact-binary inspirals (see also Ref.~\cite{Yunes:2016jcc}). It is straightforward to map 
Eq.~\eqref{Edotdip} into this generic parametrization. In our case the parameter $B$ defined in 
Refs.~\cite{Barausse:2016eii,Yunes:2016jcc} reads $B=\frac{5}{48} \zeta^2$.
The analysis of Ref.~\cite{Barausse:2016eii} shows that GW150914 sets the upper bound $|B|\lesssim 2\times 10^{-2}$, 
whereas a putative eLISA detection of a GW150914-like event with an optimal detector configuration or a combined 
eLISA-aLIGO detection set the projected bound as stringent as $|B|\lesssim 3\times10^{-9}$. In our case these bounds 
translate into 
\begin{eqnarray}
 |\zeta|&\lesssim& 0.4 \qquad \hspace{0.95cm}\text{aLIGO} \label{boundzetaLIGO}\\
 |\zeta|&\lesssim& 10^{-4}  \qquad \text{eLISA-aLIGO (projected)}  \label{boundzetaLISA}
\end{eqnarray}
for the combination $\zeta$ defined in Eq.~\eqref{zeta}. Note that the bound derived from the aLIGO detection is roughly consistent with our simplified analysis. 

Finally, we stress that if the two BHs have similar charge-to-mass ratios (which might be the case if their formation mechanisms are similar), then $\zeta\approx0$ and the dipolar emission is suppressed. In this case the zeroth-order corrections due to $G_{\rm eff}$ in Eq.~\eqref{phaseGW} would still be present [cf. Fig.~\ref{fig:Mchirp}]. In addition, the first nonvanishing radiative effect would be a quadrupolar term that will also modify the standard Newtonian quadrupole formula.

\subsection{Ringdown phase and bounds on the BH charge}
Once the two BHs merge, they form a single deformed charged and spinning BH which will relax to its final stationary (Kerr-Newman~\cite{KerrNewman}) state by
emission of GWs, EM and dark radiation. The final, ``ringdown'' stage of this process is well-described by the superposition of exponentially damped sinusoids, 
\be
\Psi(t,r) \sum_{lmn} \sim A_{lm}(r) e^{-t/\tau_{lmn}}\sin (\omega_{lmn} t)\,,
\ee
called quasinormal modes (QNMs), which are the characteristic oscillation modes\footnote{Very recently, Ref.~\cite{Cardoso:2016rao} showed that compact objects without an event horizon but with a light ring might display a ringdown signal similar to BHs even when their QNM spectrum is completely different. Such effect is due to the different boundary conditions that occur for a horizonless ultracompact object and does not play any role for charged BHs. For this reason in the rest of the paper we will refer to QNMs or to ringdown modes without distinction.} of the final BH~\cite{Kokkotas:1999bd,Ferrari:2007dd,Berti:2009kk,Konoplya:2003ii}. Here, $l$ and $m$ are angular indices describing
how radiation is distributed on the final BH's sky ($|m|\leq l$), and $n$ is an overtone index. Usually, the modes excited to larger amplitudes\footnote{As discussed in Sec.~\ref{sec:plunge}, when the BHs are highly charged the final ringdown might also depend on the $l=m=2$ Maxwell modes which might be excited to considerable amplitude. We neglect this possibility here for simplicity.} are the $(2,2,0)$ and $(3,3,0)$ gravitational modes~\cite{Buonanno:2006ui,Berti:2007fi,Barausse:2011kb}. Because the final state only depends on three parameters, the knowledge of the $(\omega_{220},\omega_{330},\tau_{220})$ triplet, for example, allows us to invert the problem and to determine the mass $M$, spin $J:=\chi M^2$ and charge $Q$ of the final BH.

To complete this program we must first know the QNMs of Kerr-Newman BHs, which has been an open problem for more than $50$ years. Fortunately, this problem was recently solved in a series of papers~\cite{Pani:2013ija,Pani:2013wsa,Zilhao:2014wqa,Mark:2014aja,Dias:2015wqa} and the QNMs of a Kerr-Newman BH can be now computed numerically.

For simplicity, here we focus on the small-charge case. The $l=m=2$ ringdown frequencies of Kerr-Newman BHs in the small-charge limit are well approximated by the following expression\footnote{We thank Aaron Zimmerman for useful correspondence on this issue and for sharing some data of Ref.~\cite{Mark:2014aja}. The small-charge results are available in Ref.~\cite{Mark:2014aja} and agree very well with the full numerical results of Ref.~\cite{Dias:2015wqa} and with the small-spin expansion of Refs.~\cite{Pani:2013ija,Pani:2013wsa}. Our fit is accurate to within $0.5\%$ in the region $j\equiv a/M\in[0,0.99]$.},
\begin{eqnarray}
 \frac{\delta \omega_{220}}{\omega_{220}}&\sim& \frac{Q^2}{M^2}\left(-0.2812-0.0243 \chi+\frac{0.3506}{(1-\chi)^{0.505}}\right)\,, \label{dwR} \\
 \frac{\delta \tau_{220}}{\tau_{220}}&\sim&-\frac{Q^2}{M^2}\left(0.1075+0.08923\chi+0.02314\chi^2+0.09443\chi^3-\frac{0.07585}{(1-\chi)^{1.2716}}\right)\,, \label{dtau}
\end{eqnarray}
where $\delta \omega_{220}\equiv \omega_{220}^{\rm KN}-\omega_{220}$, $\delta \tau_{220}\equiv \tau_{220}^{\rm KN}-\tau_{220}$, and $\omega_{220}$ and $\tau_{220}$ are the vibration frequency and damping time of a neutral, spinning BH. The latter can be found online, or computed numerically with high precision~\cite{Berti:2009kk,Berti:2005ys,ref:webpage}. 

As discussed in Appendix~\ref{app:eikonal}, there is a tight relation between the BH QNMs and the  some geodesic properties associated to the spherical photon orbits, as established in the eikonal limit~\cite{Ferrari:1984zz,Cardoso:2008bp,Yang:2012he}. Indeed, it turns out that the geodesics correspondence provides an estimate for the relations~\eqref{dwR}--\eqref{dtau}, as well as for the correction for the $(3,3,0)$ mode. Because numerical data for the $l=m=3$ modes are not available, in the rest we will estimate $\delta\omega_{330}$ from the geodesic correspondence presented in Appendix~\ref{app:eikonal}.

Let us then assume that the two dominant modes of the gravitational waveform were extracted from a GW detection, so that the two frequencies $\omega_{220}^{\rm KN}$, $\omega_{330}^{\rm KN}$
and the damping time $\tau_{220}^{\rm KN}$ were measured. In principle, using the formulas above, the mass, spin and charge of the BH could be determined precisely.
Unfortunately, detection is always done in the presence of noise, which introduces some uncertainty in the determination of the ringdown frequencies.
The proper way to handle noise is by either using Monte-Carlo simulations and a Bayesian analysis or by approximating the process through a Fisher-matrix analysis~\cite{Vallisneri:2007ev}.
A Fisher-matrix study of multi-mode ringdown is done in Ref.~\cite{Berti:2005ys}, which we follow.
For a single mode, the relevant entries are shown in the Appendix~\ref{app:fisher}.

Consider now two modes, mode 1 ($l=m=2$) with amplitude ${\cal A}_1$, frequency $\omega_1=\omega_{220}^{\rm KN}$ and damping time $\tau_1=\tau_{220}^{\rm KN}$ and mode 2 
($l=m=3$) with amplitude ${\cal A}_2$, frequency $\omega_2=\omega_{330}^{\rm KN}$ and damping time $\tau_2=\tau_{330}^{\rm KN}$. Define also the quality factor
\be
{\cal Q}_{i}=\frac{\omega_{i}\tau_{i}}{2}\,,
\ee
for $i=1,2$. For detection of multi-modes having ``orthogonal'' angular structure (i.e., the two modes are characterized by different $l,\,m$ indices), the Fisher matrix is simply an addition of matrices of different modes. In this case the errors $\sigma_{\omega_i}$ and $\sigma_{\tau_i}$ associated with frequency and damping time measurements read~\cite{Berti:2007zu}
\bea
\label{sigft-qo}
\rho \sigma_{\omega_1}&=& \f{1}{2\sqrt{2}} \left\{\f{\omega_1^3\left(3+16{\cal Q}_1^4\right)}{{\cal A}_1^2 {\cal Q}_1^7}
\left[\f{{\cal A}_1^2 {\cal Q}_1^3}{\omega_1\left(1+4{\cal Q}_1^2\right)}+\f{{\cal A}_2^2 {\cal Q}_2^3}{\omega_2\left(1+4{\cal Q}_2^2\right)}\right] \right\}^{1/2}\,,\\
\rho \sigma_{\omega_2}&=& \f{1}{2\sqrt{2}} \left\{\f{\omega_2^3\left(3+16{\cal Q}_2^4\right)}{{\cal A}_2^2 {\cal Q}_2^7}
\left[\f{{\cal A}_2^2 {\cal Q}_2^3}{\omega_2\left(1+4{\cal Q}_2^2\right)}+\f{{\cal A}_1^2 {\cal Q}_1^3}{\omega_1\left(1+4{\cal Q}_1^2\right)}\right] \right\}^{1/2}\,,\\
\rho \sigma_{\tau_1}&=&\f{4}{\pi}\left\{\f{\left(3+4{\cal Q}_1^2\right)}{{\cal A}_1^2 \omega_1 {\cal Q}_1}\left[\f{{\cal A}_1^2 {\cal Q}_1^3}{\omega_1\left(1+4{\cal Q}_1^2\right)}+
\f{{\cal A}_2^2 {\cal Q}_2^3}{\omega_2\left(1+4{\cal Q}_2^2\right)}\right] \right\}^{1/2}\,,
\eea
where $\rho$ is the signal-to-noise ratio (SNR) of the ringdown phase.

\begin{figure}[ht]
\begin{center}
\epsfig{file=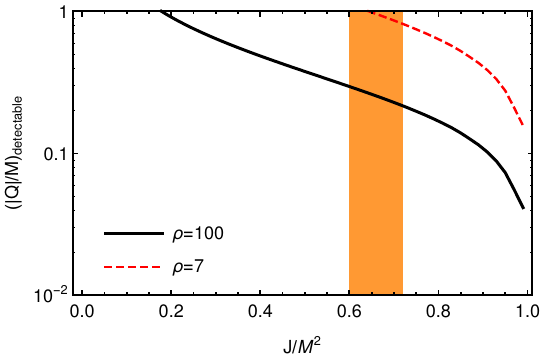,width=9cm,angle=0,clip=true}
\caption{Projected bound on the BH charge-to-mass ratio $Q/M$ as a function of the BH spin and for two different SNR of the ringdown phase, $\rho=100$ and $\rho=7$. 
We assume that the dominant mode has an amplitude approximately three times larger than the second, sub-dominant mode as is the case for many BH coalescences~\cite{Berti:2007zu}.
As a reference, the vertical band denotes the final BH spin ($J/M^2=0.67_{-0.07}^{+0.05}$) of GW150914~\cite{Abbott:2016blz} for which $\rho\approx7$ in the ringdown part~\cite{Yunes:2016jcc}. We stress that our results neglect terms of the order $Q^4/M^4$ or higher and, therefore, are only qualitative in the nearly-extremal limit.
\label{fig:sigmaQ}}
\end{center}
\end{figure}

We can convert the errors on the frequency and damping time to errors on physical quantities by using a simple propagation of errors (this procedure yields the correct analytic result in the single-mode case and we expect it to be also accurate generically, since correlations between different physical quantities are small). Specifically, we impose 
\begin{equation}
 \sigma_X=\frac{\partial X}{\partial M}\sigma_M+\frac{\partial X}{\partial \chi}\sigma_\chi+\frac{\partial X}{\partial Q}\sigma_Q\,,
\end{equation}
where $X=(\omega_1,\omega_2,\tau_1)$. It is straightforward to solve the system of three equations above for $\sigma_M$, $\sigma_\chi$ and $\sigma_Q$; this yields
\begin{eqnarray}
 \rho\sigma_M &=& f_1(\omega_1,\omega_2,\tau_1,{\cal A}_2/{\cal A}_1) \,,\\
 \rho\sigma_\chi&=& f_2(\omega_1,\omega_2,\tau_1,{\cal A}_2/{\cal A}_1)\,,\\
 \rho\sigma_Q&=& f_3(\omega_1,\omega_2,\tau_1,{\cal A}_2/{\cal A}_1)\,,
\end{eqnarray}
where $f_i$ are cumbersome analytical functions. 
Finally, we can now view $\sigma_Q$ as an upper bound on $Q$ and use it to estimate the minimum charge that can be measured by a ringdown detection with a certain SNR $\rho$. Because $\rho\sigma_Q\sim 1/Q$, the condition $\sigma_Q=Q$ gives a minimum detectable charge that scales as $1/\sqrt{\rho}$. This value is shown in Fig.~\ref{fig:sigmaQ},
for ${\cal A}_2/{\cal A}_1=1/3$, which is appropriate for a wide range of BH binaries~\cite{Berti:2007zu,Barausse:2011kb}.

Interestingly, the upper bound on $Q/M$ becomes more stringent as the final BH spin increases and it can improve by some orders of magnitude between $\chi=0$ and $\chi\sim 0.9$. The bound scales as $\sim\rho^{-1/2}$ so that it becomes more stringent for higher SNR. Thus, for a final BH with $\chi\sim 0.9$, our simplified analysis suggests that ringdown tests can set an upper constraint of the order
\begin{equation}
 \frac{|Q|}{M}\lesssim 0.1 \sqrt{\frac{100}{\rho}}\,. \label{boundQQNMs}
\end{equation}
Our analysis is valid up to ${\cal O}(Q^2/M^2)$ and should be extended to include the nearly extremal case. Nonetheless, it provides an indication for the SNR necessary to constrain the BH charge with a given SNR ratio. For example, the SNR of GW150914 is roughly $\rho\approx7$ in the ringdown part~\cite{Yunes:2016jcc}. From the spin measurement $J/M^2=0.67_{-0.07}^{+0.05}$ of GW150914~\cite{Abbott:2016blz}, our Fig.~\ref{fig:sigmaQ} suggests that the ringdown phase of GW150914 does not exclude that the final BH was nearly extremal. (We note that, for a relatively large $Q/M\sim 0.7$, our analysis neglects terms of the order $Q^4/M^4\sim0.24$, which should modify the final result by roughly a factor of $25\%$.)

\subsection{Excitation of gravitational and (hidden) EM modes in dynamical processes}\label{sec:plunge}

\begin{figure}[ht]
\begin{center}
\epsfig{file=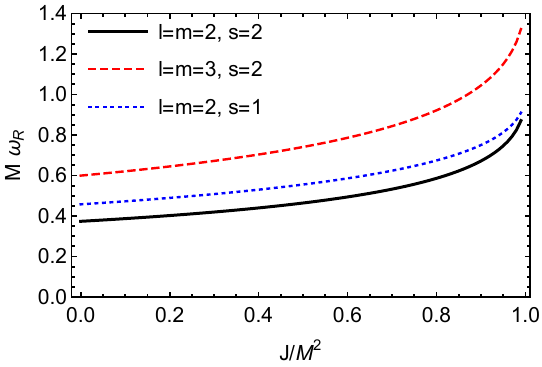,width=0.48\textwidth,angle=0,clip=true}
\epsfig{file=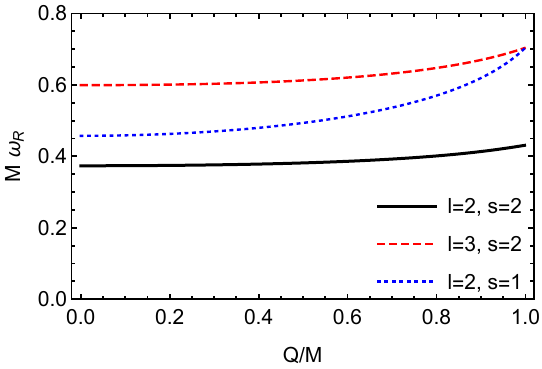,width=0.48\textwidth,angle=0,clip=true}
\caption{The frequency of the fundamental gravitational ($s=2$) and Maxwell ($s=1$) QNMs of a Kerr BH as a function of the BH spin (left panel) and of a Reissner-Nordstrom BH as a function of the BH charge (right panel). In the latter case, as $Q\to M$ the Maxwell modes with $l=2$ become isospectral to the gravitational modes with $l=3$, as a consequence of the supersymmetry of extremal Reissner-Nordstrom BHs~\cite{Onozawa:1995vu}.
\label{fig:QNMsKerr}}
\end{center}
\end{figure}

The discussion of the previous section relies on the fact that the gravitational QNMs of a Kerr-Newman BH are affected by the charge $Q$, cf. Eq.~\eqref{dwR}. However, in addition to the shift of these modes, another feature of the QNM spectrum in the presence of charge is the appearance of a new family of modes, which reduce to the standard Maxwell modes of a Reissner-Nordstrom BH when the spin vanishes.
As a reference, the fundamental gravitational and Maxwell mode of a (neutral) Kerr BH and of a static Reissner-Nordstrom BH are shown in Fig.~\ref{fig:QNMsKerr}. In the Kerr-Newman case, the modes of the Kerr BH acquire charge corrections proportional to $Q^2$ when $Q\ll M$~\cite{Mark:2014aja}, whereas the modes of the Reissner-Nordstrom BH acquire corrections proportional to $\chi$ in the small-spin case~\cite{Pani:2013ija,Pani:2013wsa}. The general case of the gravito-EM modes of a Kerr-Newman BH for arbitrary spin and charge was recently discussed in Ref.~\cite{Dias:2015wqa}.

Figure~\ref{fig:QNMsKerr} shows the well-known fact that the $l=2$ Maxwell modes of a BH are well separated from the GW modes. As we now discuss, in dynamical processes like a BH merger these modes are coupled to the gravitational sector and ---~if excited~--- they can in principle contribute to the ringdown phase of the GW. In the ringdown analysis of the previous section we have neglected such possibility and assumed that the two dominant modes were the $l=m=2$ and the $l=m=3$ gravitational modes. In this section we discuss under which conditions the extra Maxwell modes can be neglected.

For this purpose, we consider a simplified model in which a point particle (modelling a small BH) with charge $q$ and mass $\mu$ falls radially into a static BH with charge $Q$ and mass $M$. 
This model does not capture the effects of the angular momentum of the small inspiralling BH. However, in the last stages of coalescence, once the orbiting particle reaches the innermost stable circular orbit,
it will plunge into the BH. Since the ringdown is excited as the particle crosses the light ring, we suspect that
a radial infall will yield a good estimate of the effect that we're trying to study, which is the relative excitation of different QNMs.
Our model also does not capture individual spin effects. We can only hope that these effects are subdominant.

For simplicity, we consider that the charge $q$ is either a fractional electric charge $\epsilon e$ or a hidden charge $e_h$ and that the BH is charged accordingly (namely, either $Q=Q_{\rm em}$ or $Q=Q_h$). In this case, we can set either $B_{\mu\nu}$ or $F_{\mu\nu}$ to zero and the problem is effectively mapped into an electric charge $q$ plunging onto a Reissner-Nordstrom BH. The general case in which the particle and the BH have both electric and hidden charges is a simple extension of our computation.
We consider the charge and the mass of the particle to be small ($\mu\ll M$, $q\leq \mu$) so that the effect of the particle can be treated perturbatively (cf. Appendix~\ref{app:eom} for details). The metric perturbations can be analyzed through a harmonic decomposition, by separating the angular dependence of the perturbations in spherical harmonics. In the frequency domain, the EM and gravitational radiation due to a charged particle falling radially into a charged BH are described by the following coupled radial equations 
\begin{align}
&\frac{d^2\psi_g}{dr_*^2}+(\omega^2-V_g)\psi_g+I_1\psi_e=S_g, \label{eq:psig}\\
&\frac{d^2\psi_e}{dr_*^2}+(\omega^2-V_e)\psi_e+I_2\psi_g+I_3\frac{d\psi_g}{dr_*}=S_e,\label{eq:psie}
\end{align}
where $\omega$ is the frequency, and $\psi_g$ and $\psi_e$ denote the gravitational and EM master functions, respectively, and $r_*$ is the tortoise coordinate of a Reissner-Nordstrom BH, $dr/dr_*=1-2M/r+Q^2/r^2$. The potentials $V_{g,e}$, the coupling functions $I_i$, the source terms $S_{g,e}$, and a detailed derivation of the above equations are given in Appendix~\ref{app:eom}.
The source terms depend on the charge $q$, on the mass $\mu$ and on the initial Lorentz factor $\gamma$ of the particle at infinity. The coupling functions $I_i$ are proportional to the BH charge $Q$, so only when $Q=0$ the gravitational and the EM perturbations are decoupled. In the general case we expect a contamination of the EM modes into the gravitational sector and viceversa.

We employed two different methods to solve the above equations: the first is a standard Green's function technique which makes use of the solutions of the associated homogeneous systems, whereas the other is a direct integration of the full inhomogeneous system through a shooting method. We explain both procedures in Appendix~\ref{app:eom}. The two methods agree with each other within numerical accuracy, and as we explain below they reproduce earlier results in the literature.
With the solutions $\psi_{g,e}(\omega,r)$ at hand, the GW and EM energy spectra at infinity for each multipole read~\cite{Zerilli:1974ai}
\begin{eqnarray}
 \frac{dE_g}{d\omega}&=&\frac{1}{32\pi}\frac{(l+2)!}{(l-2)!}\omega^2|\psi_g(\omega,r\to\infty)|^2 \,, \label{GWflux}\\
 \frac{dE_e}{d\omega}&=&\frac{1}{2\pi}l(l+1)|\psi_e(\omega,r\to\infty)|^2\,,\label{EMflux}
\end{eqnarray}
respectively, where the wavefunctions are evaluated at spatial infinity.

\begin{figure*}[ht]
\begin{center}
\epsfig{file=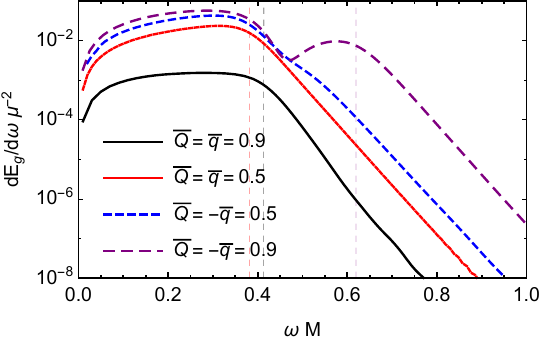,width=0.48\textwidth,angle=0,clip=true}\epsfig{file=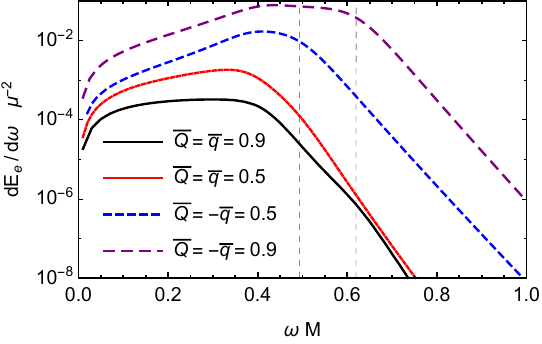,width=0.48\textwidth,angle=0,clip=true}\\
\epsfig{file=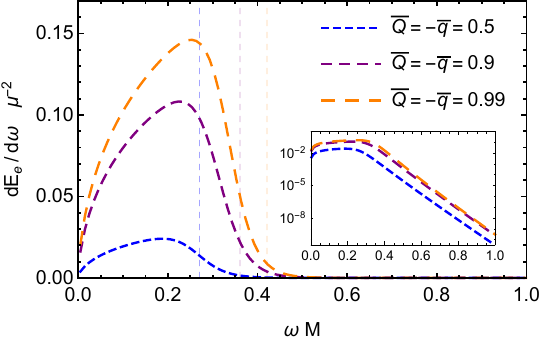,width=0.48\textwidth,angle=0,clip=true}\epsfig{file=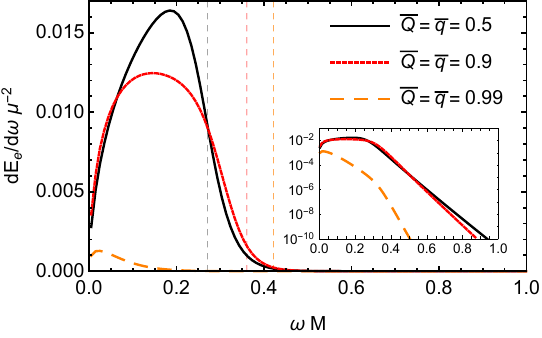,width=0.48\textwidth,angle=0,clip=true}
\caption{
Top panels: quadrupolar GW (left panel) and EM (right panel) energy spectra for a particle with charge plunging radially on a RN BH. The vertical lines denote the $l=2$ fundamental QNMs of the RN BH with $Q/M=0.5,0.9$, cf. Table~\ref{tab:modes}.
The rightmost vertical line in the left panel denotes the $l=2$ EM mode with $Q/M=0.9$, which also excited in the \emph{gravitational} spectrum.
Bottom panels: dipolar EM energy spectra for the case of opposite (left panel) and equal (right panel) charges. The insets show the same plot in a logarithmic vertical scale. In all cases the initial Lorentz factor of the point charge is $\gamma=1.001$. The vertical lines denote the $l=1$ fundamental EM QNMs of the RN BH with $Q/M=0.5,0.9,0.99$, cf. Table~\ref{tab:modes}.
In the legend of all panels $\bar{Q}:=Q/M$ and $\bar{q}:=q/\mu$.
\label{fig:spectrum}}
\end{center}
\end{figure*}

The energy spectra defined above are shown in Fig.~\ref{fig:spectrum} for some representative cases. The most salient (and more relevant for this discussion) feature
of the GW spectrum (top left panel) is that it has a cutoff at roughly the lowest QNM frequency of the central BH. A similar cut is also present for the EM energy spectrum with $l=2$ (top right panel) and with $l=1$ (bottom panels). The vertical lines in the panels of Fig.~\ref{fig:spectrum} denote the fundamental QNMs, which are presented in Table~\ref{tab:modes} for completeness. The largest contribution to the EM energy spectrum comes from the dipole (bottom panels) which, for sufficiently large values of $|Q|\sim|q|$, is even larger than the GW energy flux.

 \begin{table}[tbp]
 \centering
 \begin{tabular}{cccc}
 \hline\hline
  $Q/M$ & $(s,l)$ & $\omega_R M$ & $-\omega_I M$ \\
\hline
  $0.99$ & $(2,2)$ & $0.4035$ & $0.2570$ \\
  $0.99$ & $(1,2)$ & $0.6787$ & $0.2675$ \\
  $0.99$ & $(1,1)$ & $0.4214$ & $0.0871$ \\
\hline
  $0.9$ & $(2,2)$ & $0.4136$ & $0.0883$ \\
  $0.9$ & $(1,2)$ & $0.6194$ & $0.0976$ \\
  $0.9$ & $(1,1)$ & $0.3608$ & $0.0974$ \\
\hline
  $0.5$ & $(2,2)$ & $0.3817$ & $0.0896$ \\
  $0.5$ & $(1,2)$ & $0.4937$ & $0.0972$ \\
  $0.5$ & $(1,1)$ & $0.2707$ & $0.0951$ \\
  \hline \hline
 \end{tabular}
 \caption{Fundamental QNMs of a Reissner-Nordstrom BH computed with continued fractions~\cite{Berti:2009kk} for different values of the BH charge. Gravitational-led and EM-led modes are denoted by $s=2$ and $s=1$, respectively.}\label{tab:modes}
\end{table}

We are now in a position to consider our initial question on the excitation of the EM modes in dynamical processes. Figure~\ref{fig:spectrum} shows that the flux for $Q=-q=0.9 M$ displays two peaks: the first one corresponds to the excitation of the $l=2$ gravitational mode, whereas the second peak corresponds to the $l=2$ EM mode. The latter is excited due to the coupling between gravitational and EM perturbations [cf. Eqs.~\eqref{eq:psig}--\eqref{eq:psie}], which is a feature of charged BH spacetimes~\cite{Johnston:1973cd,Johnston:1974vf}. In other words, the GWs emitted in the process contain information about the EM modes of central BH, not only about its gravitational modes.

The ratio between the flux at the two peaks depends on the charge and on the initial Lorentz factor $\gamma$ of the particle, and is well fitted by
\be
\mathcal{R}\equiv \frac{dE_g/d\omega|_{\omega=\omega_e}}{dE_g/d\omega|_{\omega=\omega_g}}\sim a+b\left(\frac{Q}{M}\right)^2+c\left(\frac{Q}{M}\right)^3, \label{ratio}
\ee
where $\omega_{e,g}$ are the frequencies at the peaks, and the coefficients $(a,b,c)$ depend on $\gamma$. In Appendix~\ref{app:eom} we show some supplemental results about the energy fluxes emitted in the process. In particular, we show the ratio~\eqref{ratio} for some values of $\gamma$ [cf. Fig.~\ref{fig:ratio_f} in Appendix~\ref{app:eom}]. 
%
Thus, the relative amplitude of the EM peak is larger when $\gamma\sim 1$, at least for $\gamma\lesssim1.2$. Furthermore, ${\cal R}$ is nonnegligible only when the BH is near extremality and only when $Qq<0$.

\subsection{Emission of (hidden) EM radiation in a binary BH merger}

\begin{figure}[ht]
\begin{center}
\epsfig{file=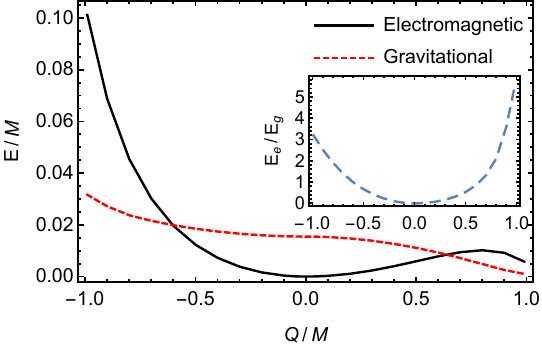,width=7.8cm,angle=0,clip=true}
\caption{
Total energy radiated to infinity, summing the gravitational multipoles up to $l=3$ and the EM multipoles up to $l=2$. We consider $\gamma=1.1$ and the charge-to-mass ratio of the particle $q/\mu=|Q|/M$. 
\label{fig:total_energy}}
\end{center}
\end{figure}

Let us now discuss the total energy radiated in the collision. This is shown in Fig.~\ref{fig:total_energy} for the representative case of an initial Lorentz factor $\gamma=1.1$ of the point particle. 
An interesting aspect of this figure is that in the extremal limit, $Q=M$, both the EM and gravitational radiation are suppressed.
The reason is that in this limit the gravitational attraction cancels the EM repulsion, leading to constant dipole and quadrupole moments
(there is in fact a static solution of the field equations describing two maximally charged BHs in equilibrium, the Majumdar-Papapetrou solution~\cite{Majumdar:1947eu,Papapetrou:1945eu}).

For collisions from rest ($\gamma\approx 1$), assuming $Qq>0$, we obtain 	
\begin{align}
\frac{E_{e}}{M}&\approx0.021\frac{q^2\mu^2}{M^4}\,,\\
\frac{E_g}{M}&\approx0.01\frac{\mu^2}{M^2}-0.015\frac{Q^2\mu^2}{M^4} \,.
\end{align}
This result can be compared with flat-space estimates using simple quadrupole and dipole Larmor formulae for the GW and EM emission, respectively.
For the GW flux we get
\be
\frac{dE_g}{dt}=\frac{2}{15}\frac{\partial^3}{\partial t^3}|D_{rr}|^2\,,\qquad D_{rr}=\mu r(t)^2\,.
\ee
Using energy conservation $\dot{r}=\sqrt{2M/r}$ for infalls from rest, and we obtain
\be
E_g=\int_{2M}^{\infty}\frac{1}{\dot{r}}\frac{dE}{dt}=\frac{2\mu^2}{105 M}\sim 0.019\frac{\mu^2}{M}\,.
\ee
The same procedure for the flat-space Larmor formula gives
\be
E_e=\int_{2M}^{\infty}\frac{1}{\dot{r}}\frac{2q^2\ddot{r}^2}{3}=\frac{q^2}{30 M}\sim 0.033 \frac{q^2}{M}\,.
\ee
Thus, a flat-space Newtonian calculation agrees with our numerical results to within a factor two.

An interesting quantity is the ratio $E_e/E_g$ which is shown as a function of $Q$ in the inset of Fig.~\ref{fig:total_energy}. From the above relations, in the small-$Q$ limit we obtain
\begin{equation}
 \frac{E_e}{E_g}\approx \frac{Q^2}{M^2}\times\left\{
\begin{array}{ll}
1.8\,, &~{\rm for}~ \gamma=1.1,\\	
2.0\,, &~{\rm for}~ \gamma\approx 1,\\	
\end{array}\right. \label{EeoEg}
\end{equation}
which is of the order of $Q^2/M^2$ as expected.

Up to now, our results are formally valid only in a perturbative scheme; however, decades of work shows that even fully nonlinear results for equal-mass BHs can be recovered with point-particle calculations if one replaces $\mu$ with the reduced mass of the system~\cite{Cardoso:2014uka,Tiec:2014lba}. This substitution would immediately recover (within a factor of two) the results in Ref.~\cite{Zilhao:2012gp} for the gravitational radiation produced during the head-on collisions of two equal-mass, equal-charge BHs. However, our point particle results {\it overestimate} the amount of EM radiation during that same process. The reason is that, as we remarked earlier, for two equal mass-to-charge ratio objects, dipole emission is suppressed and one only gets quadrupole emission. In fact, our $l=2$ results for the EM channel agree with the numbers reported in Ref.~\cite{Zilhao:2012gp} once the extrapolation to equal-mass is done.

In a binary BH coalescence, the total energy loss in GWs can be enormous. For example, the GW luminosity of GW150914 was $dE_g/dt\approx 3.6\times 10^{56}{\rm erg/s}$~\cite{Abbott:2016blz}. Equation~\eqref{EeoEg} shows that, even for BHs with small $Q$, the EM luminosity can still be very large. In this model the EM luminosity of a GW150914-like event is roughly
\begin{equation}
 \frac{d E_e}{dt}\sim 7\times 10^{56} \left(\frac{Q}{M}\right)^2\,{\rm erg/s}\,. \label{EMluminosity}
\end{equation}
Even for weakly charged BHs with $Q\sim 10^{-4}M$, a GW150914-like event would produce an EM luminosity $dE_e/dt\approx 10^{48}{\rm erg/s}$, comparable to the luminosity of the weakest gamma-ray bursts~\cite{2012ApJ...756..190Z}.
	
Nonetheless, the spectra presented in Fig.~\ref{fig:spectrum} show that there exists a cutoff frequency associated to the fundamental EM QNM of the final BH. This mode has a typical frequency of the order of [cf. Table~\ref{tab:modes}]
\begin{equation}
 f_{\rm QNM}\approx 300 \left(\frac{60 M_\odot}{M}\right) \,{\rm Hz}\,,\label{fQNM}
\end{equation}
(the precise value depends on the spin and on the charge [cf. Fig.~\ref{fig:QNMsKerr}]). If the BH is electrically charged, the energy is released in GWs and ordinary photons with frequency $f\lesssim f_{\rm QNM}$. Low-frequency photons are absorbed by the interstellar medium if their frequency is smaller than the plasma frequency
\begin{equation}
f_{\rm plasma}^{\rm em}=(2\pi)^{-1}\sqrt{\frac{4\pi n_e e^2}{m_e}}\sim 10^4\sqrt{\frac{n_e}{1\,{\rm cm}^{-3}}}\, {\rm Hz}\,,
\end{equation}
where $n_e$ is the electron number density. Photons with frequency $f<f_{\rm plasma}^{\rm em}$ do not propagate in the plasma. In the interstellar medium $n_e\approx 1\,{\rm cm}^{-3}$, so most of the EM energy released in the process is absorbed by the plasma.

On the other hand, if the BH is charged under a hidden $U(1)$ charge, a sizeable fraction of the luminosity is released in dark photons. The latter do not interact with ordinary electrons, but would nevertheless interact with hidden plasma, whose typical frequency reads
\begin{equation}
f_{\rm plasma}^h=(2\pi)^{-1}\sqrt{\frac{4\pi n_{e_h}  e_h^2}{m}}\sim 250\left(\frac{\rho_{\rm DM}}{0.4\,{\rm GeV}/{\rm cm^3}}\right)^{1/2}\left(\frac{e_h/m}{10^{-3} e/m_e}\right)\, {\rm Hz}\,, \label{fplasmahidden}
\end{equation}
where we estimated $n_{e_h}\sim \rho_{\rm DM}/m$ and we normalized the DM density $\rho_{\rm DM}$ to its typical local value, $\rho_{\rm DM}\sim 0.4\,{\rm GeV}/{\rm cm^3}$. Comparison between Eq.~\eqref{fplasmahidden} and Eq.~\eqref{fQNM} shows that the frequency of hidden photons is larger than a typical hidden-plasma frequency whenever 
\begin{equation}
 \epsilon_h:=\frac{e_h}{e}\ll 2.4 \left(\frac{m}{{\rm GeV}}\right) \left(\frac{60 M_\odot}{M}\right)\left(\frac{0.4\,{\rm GeV/cm^3}}{\rho_{\rm DM}}\right)^{1/2}\,,\label{boundplasma}
\end{equation}
The above threshold line is shown in the right panel Fig.~\ref{fig:bounds}. In the region below the dark red dot-dashed line, the hidden photons emitted during the ringdown of a charged BH with $M\sim 60M_\odot$ propagate freely in a hidden plasma of density $\rho_{\rm DM}\sim 0.4\,{\rm GeV/cm^3}$. Interestingly, the phenomenologically-viable region in the right panel Fig.~\ref{fig:bounds} lies below the threshold line, so in this region hidden photons are expected to propagate freely.

\section{Discussion and final remarks} \label{sec:discusscf}
We have shown that, in models of minicharged DM and dark radiation, astrophysical BHs can have large (electric and/or hidden) charge and are uniquely described by the Kerr-Newman metric. In these models the standard arguments that prevent astrophysical BHs to have some electric charge within Einstein-Maxwell theory can be circumvented and, in particular, nearly-extremal BHs with $Q\sim M$ are also phenomenologically available.

Charged BHs in these scenarios are interesting GW sources. The GW signal from the coalescence of two charged BHs contains a wealth of information about the properties of the system. In particular, we have shown that the inspiral and the post-merger ringdown stages provide complementary information. The initial inspiral can constrain a combination of the initial BH charges [cf. Eq.~\eqref{zeta}] and a rescaling of Newton's constant, whereas ringdown tests are only sensitive to the final BH total charge.

\emph{Combined} inspiral and ringdown tests might also be performed provided the energy lost in GWs and EM waves during the coalescence is known. Therefore, it would be very interesting to perform fully numerical simulations of charged-BH binary systems close to coalescence (extending the work of Ref.~\cite{Zilhao:2012gp}) and to estimate the mass, spin and charge of the final Kerr-Newman BH formed after the merger.
The combined information from the inspiral and ringdown phases (together with estimates of the mass and charge loss during the merger phase) can be used to disentangle part of the degeneracy appearing in the dipole formula. Such information would be crucial for cross checks similar to those performed for the masses and spins of GW150914~\cite{Abbott:2016blz,TheLIGOScientific:2016src}. A more detailed analysis in this direction is left for future work.

Likewise, our analysis of the upper bounds derived from ringdown detections can be improved in several ways, for example by considering multiple detections, multiple modes, and a more sophisticated statistical analysis. It is also reasonable to expect that GW150914 was not a statistical fluctuation and that even louder GW events (with higher SNR in the ringdown phase) might be detected in the near future, when second-generation detectors will reach their design sensitivity. In this case our results suggest that an analysis of the entire GW signals can provide stringent constraints on the charge of BHs in minicharged DM models; we hope that this exciting prospect will motivate further studies on this topic.

BH-based tests provide a unique opportunity to constrain the hidden coupling $e_h$, which is otherwise challenging to probe with EM-based tests. An interesting prospect in this direction is the burst of low-frequency dark photons emitted during the merger [cf. Eq.~\eqref{EMluminosity}]. As we have shown, these dark photons are not absorbed by the DM plasma and their luminosity can be extremely high. In some models of minicharged DM, dark photons are coupled to ordinary photons through a kinetic mixing term~\cite{Holdom:1985ag,Burrage:2009yz,Foot:2014uba} proportional to $\sim \sin\theta$ (where $\tan\theta:=\epsilon/\epsilon_h$), so that conversion of dark photons to ordinary photons might occur when $\epsilon\neq0$. The frequency of dark photons emitted in BH mergers are typically smaller than the kilohertz, and therefore next-to-impossible to detect with ordinary telescopes. 
However, there might be mechanisms in which dark photons can convert to higher-frequency ordinary photons and to ordinary fermions. This conversion might provide an exotic EM counterpart of BH mergers and might leave a detectable signal in current experiments~\cite{Feng:2015hja,Feng:2016ijc}.
Futhermore, it is in principle possible that the (gigantic) burst of dark photons affects nearby (hidden-) charged stars, via the same mechanisms we described. In such a case, a passing burst of hidden radiation would cause nearby stars
to oscillate, with r.m.s fluctuations that could be measurable and extracted from asteroseismology studies, in a phenomena similar to that described recently for GWs~\cite{Lopes:2015pca,Lopes:2014dba}.

We have focused on models of massless dark photons. Massive dark photons are an appealing candidate to explain the muon $g-2$ discrepancy~\cite{Davoudiasl:2012ig}. If ultralight, these bosons are known to turn spinning BHs unstable~\cite{Pani:2012vp,Pani:2012bp,Witek:2012tr} due to a superradiant instability (cf. Ref.~\cite{Brito:2015oca} for an overreview). A more rigorous analysis of this instability is a further interesting application of BH physics in the context of dark-radiation models.

An important open question concerns the formation of charged BHs in these scenarios. Accretion of charged DM particles is a natural charging mechanism. Charged BHs might also form in the gravitational collapse of charged compact stars, the latter might acquire (electric or hidden) charge by DM capture in their interior. To the best of our knowledge there are no studies on the DM accretion by BHs in models of minicharged DM and ---~in light of our results~--- it would be very interesting to fill this gap (see Refs.~\cite{Banks:2014rsa,Fischler:2014jda} for some related work). GW-based bounds on the BH charge might be combined to realistic accretion models to constrain the parameter space of minicharged DM models.

%

\begin{acknowledgments}
We are indebted to Leonardo Gualtieri for many useful discussions during the development of this project and to Kent Yagi and Nico Yunes for valuable comments on the draft. We thank Aaron Zimmerman for useful correspondence and for sharing some data of Ref.~\cite{Mark:2014aja} with us.
We thank \O yvind Christiansen for pointing a few typos and errors in the published version, which are now corrected 
on the arxiv version, and described in an erratum in the journal. These do not affect our results at qualitative level.
V.C. acknowledges financial support provided under the European Union's H2020 ERC Consolidator Grant ``Matter and strong-field gravity: New frontiers in Einstein's theory'' grant agreement no. MaGRaTh--646597.
C.M. acknowledges financial support from Conselho Nacional de Desenvolvimento Cient\'ifico e Tecnol\'ogico through Grant No.~232804/2014-1.
Research at Perimeter Institute is supported by the Government of Canada through Industry Canada and by the Province of Ontario through the Ministry of Economic Development $\&$ Innovation.
This work was supported by the H2020-MSCA-RISE-2015 Grant No. StronGrHEP-690904 and by FCT-Portugal through the project IF/00293/2013.
%
\end{acknowledgments}

\appendix

\section{Quasinormal modes of Kerr-Newman BHs from the geodesic correspondence}\label{app:eikonal}

In this appendix we discuss the correspondence between the BH QNMs and the properties of spherical photon orbits~\cite{Ferrari:1984zz,Cardoso:2008bp,Yang:2012he}. In the static case, the real part of the QNM frequency is related to the azimuthal orbital frequency, whereas the imaginary part of the frequency corresponds to the Lyapunov exponent of the orbit~\cite{Cardoso:2008bp}. In the rotating case the relation between modes with generic $(l,m)$ and some geodesic properties is more involved~\cite{Yang:2012he}. For simplicity, here we focus on the $l=m$ case in which the analysis is remarkably straightforward, since these modes are associated only with equatorial motion~\cite{Yang:2012he}.

Let us start with the stationary and axisymmetric line element 
\begin{equation}
 ds^2=g_{tt}dt^2+g_{rr}dr^2+g_{\theta\theta}d\theta^2+2g_{t\phi}dtd\phi+g_{\phi\phi}d\phi^2\,,
\end{equation}
where all metric coefficients are functions of $r$ and $\theta$ only. The radial motion of null particles on the equatorial plane is governed by 
\begin{equation}
 \dot{r}^2=V:=\frac{E^2 g_{\phi\phi}+2 E L g_{t\phi}+g_{tt} L^2}{g_{rr} g_{t\phi}^2-g_{tt} g_{rr} g_{\phi\phi}}\,,
\end{equation}
where $E$ and $L$ are the (conserved) specific energy and angular momentum of the geodesic, the metric coefficients are evaluated on the equatorial plane, and a dot denotes derivative with respect to the affine parameter of the null geodesic. The light ring and the corresponding ratio $L/E$ are defined by $V=0=V'$, where a prime denotes a radial derivative. The orbital frequency of the light ring is simply the azimuthal frequency,
\begin{equation}
 \Omega=\frac{-g_{t\phi}'+\sqrt{{g_{t\phi}'}^2-g_{tt}' g_{\phi\phi}'}}{g_{\phi\phi}'}
\end{equation}
evaluated at the light-ring location. Other relevant orbital frequencies are
\begin{eqnarray}
  \Omega_\theta &=& \frac{g_{tt}+\Omega g_{t\phi}}{\sqrt{2 g_{\theta\theta}}}\sqrt{\frac{\partial^2 U}{\partial \theta^2}}\,, \label{Omegatheta} \\
  \Omega_{\rm pre} &=& \Omega-\Omega_\theta\,, \label{Omegapre}
\end{eqnarray}
where $U(r,\theta)=g^{tt}-2(L/E)g^{t\phi}+(L/E)^2 g^{\phi\phi}$ and the expressions above are evaluated at the light-ring location on the equatorial plane. $\Omega_\theta$ represents the frequencies of small oscillations around quasicircular equatorial orbits along the angular direction, respectively, whereas $\Omega_{\rm pre}$ is the precession frequency of the orbital plane.

Reference~\cite{Yang:2012he} shows that the QNM frequency in the eikonal limit reads
\begin{equation}
 \omega_R \sim (l+1/2)\Omega_\theta+m\Omega_{\rm pre} \sim l(\Omega_\theta+\Omega_{\rm pre})\,,
\end{equation}
where in the last step we used $l=m\gg1$. Interestingly, in the $l=m$ limit the above expression coincides with that derived for static spacetimes in Ref.~\cite{Cardoso:2008bp}, i.e. $\omega_R\sim l\Omega$.

Although not relevant for our analysis, for completeness we discuss the geodesic correspondence of the damping time of the modes. In the eikonal limit the latter is related to the Lyapunov coefficient of the orbit~\cite{Cardoso:2008bp}
\begin{equation}
 \lambda = -\frac{1}{\dot t}\sqrt{\frac{V''}{2}}\,, \qquad \dot t=-\frac{E^2 g_{\phi\phi}+L g_{t\phi}}{g_{t\phi}^2-g_{tt}g_{\phi\phi}}
\end{equation}
again evaluated at the light-ring location on the equatorial plane. Thus, for $l=m\gg1$ the complex QNM can be written as~\cite{Ferrari:1984zz,Cardoso:2008bp,Yang:2012he} 
\begin{equation}
 \omega_R+i\omega_I \sim \Omega l-i(n+1/2)|\lambda|\,, \label{eikonal}
\end{equation}
where $n$ is the overtone number. Note that the above expression formally coincides with that obtained in Refs.~\cite{Cardoso:2008bp} for static spacetimes and it extends the results of Ref.~\cite{Ferrari:1984zz} which are valid only for slowly-rotating BHs. A more involved result for QNMs with generic $(l,m)$ is derived in Ref.~\cite{Yang:2012he}.

\begin{figure*}[ht]%
\begin{center}
\includegraphics[width=0.49\textwidth]{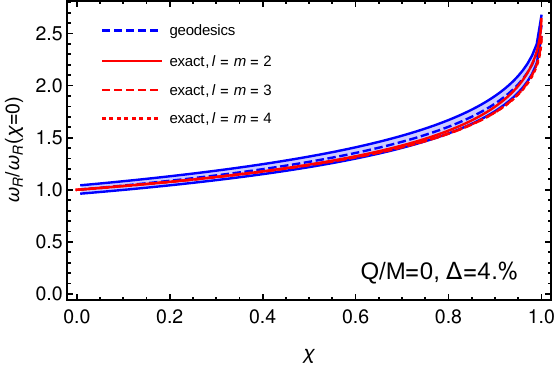}
\includegraphics[width=0.49 \textwidth]{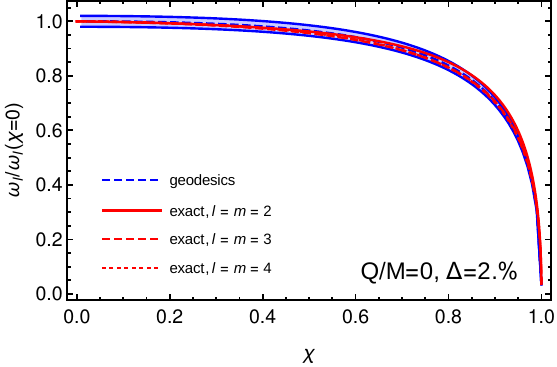}\\
\includegraphics[width=0.49 \textwidth]{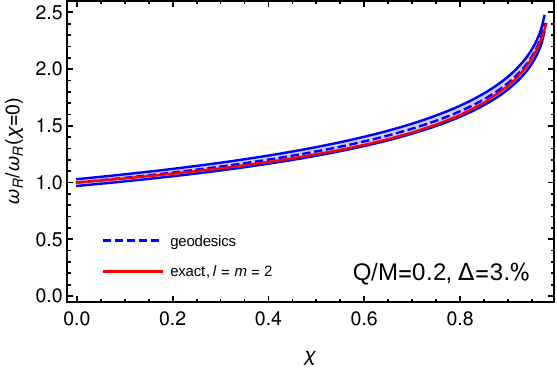}
\includegraphics[width=0.49 \textwidth]{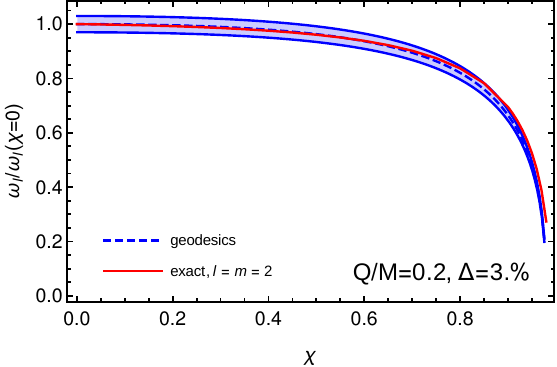}
\caption{Comparison between the exact gravitational QNMs of a Kerr-Newman BH and those obtained from a geodesic analysis [cf. Eq.~\eqref{eikonal}] which is formally valid for $l=m\gg1$. We show the ratio between the real (left panels) and the imaginary (right panels) part of the QNM frequency as a function of the BH spin and normalized by their value in the nonspinning case.  Top panels show the case of a Kerr BH, bottom panels show the case of a Kerr-Newman BH with $Q=0.2M$ (in the latter case the exact QNMs are available only for $l=2$~\cite{Mark:2014aja}). The blue band brackets a percentage variation $\Delta$ relative to the geodesic prediction.
With the adopted normalization, the curves for $l=m=3$ and $l=m=4$ modes of a Kerr BH are almost identical.}%
\label{fig:eikonal}%
\end{center}
\end{figure*}

Strictly speaking, the geodesic prediction~\eqref{eikonal} should only be valid in the eikonal limit, i.e. when $l\gg1$. However, in Fig.~\ref{fig:eikonal} we show that the analytical result~\eqref{eikonal} agrees remarkably well with the exact numerical results for the QNMs of a Kerr-Newman BH even when $l=m=2$. Relative errors are always smaller then $\approx 4\%$ for any spin, both in the neutral case (top panels of Fig.~\eqref{fig:eikonal}) and in the Kerr-Newman case with $Q/M=0.2$ (bottom panels of Fig.~\eqref{fig:eikonal}). In the latter case the exact results are only available for $l=2$~\cite{Mark:2014aja}, but their deviation from the geodesic predition is always smaller than $3\%$.

In the main text, we have used this striking agreement to estimate the $l=m=3$ modes of a weakly-charged Kerr-Newman BH. Note that the deviations from the geodesic predictions are likely smaller than the observational errors on these modes, therefore this approximation should not affect our analysis significantly.

\section{Fisher matrix analysis}\label{app:fisher}
We follow Ref.~\cite{Berti:2005ys} for the analysis of uncertainties associated with measurements of
ringdown waveforms in noise. We assume that the GW signal during the ringdown phase
can be expressed as a linear superposition of exponentially decaying
sinusoids, the QNMs of the spacetime.
The gauge invariant waveforms are then given by
\be 
h_++\ii h_{\times}=\frac{M}{r}\sum _{lmn}
{\cal A}_{lmn}
e^{\ii(\omega_{lmn} t+\phi_{lmn})}e^{-t/\ta} \Slm\,. \label{QNMexp}
\ee
In this expansion the spheroidal functions $\Slm=S_{lm}(a\om)$ are evaluated at the (complex) QNM frequencies, so they are complex numbers (henceforth we drop the angular dependence on the $\Slm$).

The waveform measured at a detector is given by
\be
h = h_{+} F_+(\theta_S,\phi_S,\psi_S) + h_{\times}
F_\times(\theta_S,\phi_S,\psi_S) \,,
\label{detectwave}
\ee
where $F_{+,\times}$ are pattern functions that depend on the orientation
of the detector and the direction of the source, namely
%
\begin{subequations}
\bea
F_+(\theta_S,\phi_S,\psi_S) &=& \frac{1}{2}(1 + \cos^2 \theta_S) \cos 2\phi_S \cos
2\psi_S
-\cos \theta_S \sin 2\phi_S \sin 2\psi_S\;, \\
F_{\times}(\theta_S,\phi_S, \psi_S) &=& \frac{1}{2}(1 + \cos^2 \theta_S) \cos
2\phi_S \sin 2\psi_S
+ \cos \theta_S \sin 2\phi_S \cos 2\psi_S\;.
\label{pattern}
\eea
\end{subequations}
%

We will follow the prescription outlined in Ref.~\cite{Berti:2005ys} to compute the SNR $\rho$. We assume large quality factor ${\cal Q}_{lmn}$ and average the source over sky position and over detector and BH orientations, making use of the angle averages: $\langle F_+^2\rangle=\langle F_\times^2\rangle=1/5$, $\langle F_+F_\times\rangle=0$, and
$\langle |\Slm|^2 \rangle=1/4\pi$.

With a given noise spectral density for the detector,
$S_h(f)$, one defines the inner product between two signals $h_1(t)$
and $h_2(t)$ by
\begin{equation}
(h_1|h_2) \equiv 2 \int_0^{\infty} \frac{ {\tilde{h}_1}^*\tilde{h}_2 +
{\tilde{h}_2}^*\tilde{h}_1 }{S_h(f)}df \,,
\label{innerproduct}
\end{equation}
where $\tilde{h}_1(f)$ and $\tilde{h}_2(f)$ are the Fourier transforms
of the respective gravitational waveforms $h(t)$.  The components of
the Fisher matrix $\Gamma_{ab}$ are then given by
\begin{equation}
\Gamma_{ab} \equiv \left( \frac{\partial h}{\partial\theta^a} \mid
\frac{\partial
h}{\partial\theta^b} \right) \,,
\label{fisher}
\end{equation}
where $\theta^a$ are the source parameters.
In the limit of large SNR, if the noise is stationary and Gaussian,
the probability that the GW signal $s(t)$ is characterized by a given
set of values of the source parameters $\theta^a$ is
\be
p(\mbox{\boldmath$\theta$}|s)=p^{(0)}(\mbox{\boldmath$\theta$})
\exp\left[-\frac{1}{2}\Gamma_{ab}\delta \theta^a \delta \theta^b\right]\,.
\ee
where $\delta \theta^a = \theta^a - {\hat \theta}^a$, and
$p^{(0)}(\mbox{\boldmath$\theta$})$ represents the distribution of prior
information.
An estimate of the rms error, $\Delta\theta^a = (\langle (\delta \theta^a)^2
\rangle )^{1/2}$, in measuring the
parameter $\theta^a$ can then be calculated, in the limit of large
SNR, by taking the square root of the diagonal elements of the inverse
of the Fisher matrix,
\begin{equation}
\Delta\theta^a = \sqrt{\Sigma^{aa}} \,, \qquad  \Sigma = \Gamma^{-1} \,.
\label{errors}
\end{equation}

The Fisher matrix components in the parameter basis of $(A^{+},\, A^{\times},\, \ph^{+},\, \ph^{\times},\, \flm,\, {\cal Q}_{lmn})$, were computed and presented in Ref.~\cite{Berti:2005ys} [here $\flm:=\omega_{lmn}/(2\pi)$]. In the large $\Qlm$ limit, they read 
\allowdisplaybreaks{
\begin{subequations}
\bea
\Gamma_{A^+A^+} &=&\Gamma_{A^\times A^\times}= \f{\gamma}{A^2} \left (1+4{\cal Q}_{lmn}^2 \right) \,, \nonumber\\
\Gamma_{A^+ A^\times}  &=& \Gamma_{A^+\ph^{\times}} =\Gamma_{A^\times\ph^{+}}= \Gamma_{A^+ \ph^+} =\Gamma_{A^\times\ph^\times} = 0 \,,\nonumber\\
\Gamma_{A^+ \flm} &=& - \f{\gamma }{2 A \flm} \left (1+4{\cal Q}_{lmn}^2  \right) \cos \psi \,,\nonumber\\
\Gamma_{A^\times \flm} &=& - \f{\gamma }{2 A \flm}\left (1+4{\cal Q}_{lmn}^2  \right) \sin \psi \,,\nonumber\\
\Gamma_{A^+ {\cal Q}_{lmn}} &=&  \f{\gamma}{2 A \Qlm}\f{1}{1+4{\cal Q}_{lmn}^2}\left [(1+4{\cal Q}_{lmn}^2)^2  \right ] \cos \psi \,,\nonumber\\
\Gamma_{A^\times {\cal Q}_{lmn}} &=&  \f{\gamma}{2 A {\cal Q}_{lmn}}\f{1}{1+4{\cal Q}_{lmn}^2}(1+4{\cal Q}_{lmn}^2)^2 \sin \psi \,,\nonumber\\
\Gamma_{\ph^+ \ph^+ } &=& \gamma \left (1+4{\cal Q}_{lmn}^2 \right)\cos^2 \psi\,,\nonumber\\
\Gamma_{\ph^\times \ph^\times } &=& \gamma \left (1+4{\cal Q}_{lmn}^2 \right) \sin^2 \psi\,,\nonumber\\
\Gamma_{\ph^+ \ph^\times } &=& \Gamma_{\ph^+ \flm} =\Gamma_{\ph^\times \flm}=\Gamma_{\ph^+ \Qlm} =\Gamma_{\ph^\times {\cal Q}_{lmn}} = 0 \,,\nonumber\\
\Gamma_{\flm \flm} &=& \f{\gamma}{2\flm^2} (1+4{\cal Q}_{lmn}^2)^2 \,,\nonumber\\
\Gamma_{\flm {\cal Q}_{lmn}} &=& -\f{\gamma}{2\flm {\cal Q}_{lmn}}\f{1}{1+4{\cal Q}_{lmn}^2}\left [ (1+4{\cal Q}_{lmn}^2)^2 \right ] \,,\nonumber\\
\Gamma_{{\cal Q}_{lmn}{\cal Q}_{lmn}} &=& \f{\gamma}{2 {\cal Q}_{lmn}^2}\f{1}{(1+4{\cal Q}_{lmn}^2)^2}(1+4{\cal Q}_{lmn}^2)^3  \,.\nonumber
\eea
\label{fishermatrix}
\end{subequations}
}
Here, $A^2 = (A^+)^2 + (A^\times)^2$, $\cos \psi \equiv A^+/A$, $\sin \psi \equiv A^\times /A$, and
\be
\gamma = \f{A^2 {\cal Q}_{lmn} }{40\pi^2 \flm (1+4{\cal Q}_{lmn}^2)} \,.
\ee
We also have $\rho^2 = \gamma (1+4{\cal Q}_{lmn}^2-\beta)$. Finally, the transformation from the
$(\flm,\, {\cal Q}_{lmn})$ basis to the $(Q, \,\chi)$ basis reads
\be
\Gamma_{k\,j}=\flm'  \Gamma_{k\,\flm} + {\cal Q}_{lmn}' \Gamma_{k\,{\cal Q}_{lmn}} \,,
\ee
for any index $k\neq j$, and where $\flm' := d\flm/d\chi$ and ${\cal Q}_{lmn}' := d{\cal Q}_{lmn}/d\chi$.

\section{Technical details on the radial infall of a point charge into a charged BH}\label{app:eom}
In this appendix we give some details about the derivation and integration of the coupled system~\eqref{eq:psig}--\eqref{eq:psie}. We follow the procedure outlined in
Ref.~\cite{Zerilli:1974ai}, but correcting typos and possible errors. Our integration technique is fully consistent, and has been validated by two independent codes,
as well as with previous results in the literature for uncharged BHs. One of the codes, written in {\it Mathematica}, is freely available online~\cite{files_online}.

\subsection{EM and metric perturbations}
The spacetime metric due to a point charge falling into a charged BH can be written as $g_{ab}^{(0)}+h_{ab}$, where $h_{ab}\ll g_{ab}^{(0)}$ is a small perturbation to the background Reissner-Nordstrom geometry, 
\be
g_{ab}^{(0)}=-e^{v}dt^2+e^{-v}dr^2+r^2d\theta^2+r^2\sin^2\theta d\phi^2\,,
\ee
where $e^{v}=1-2M/r+Q^2/r^2$. Here we consider the Regge-Wheeler decomposition of $h_{ab}$. In the case of a radially falling particle, the metric perturbations have even (polar) parity and $h_{ab}$ can be written as
\be
h_{ab}=
\l(
\begin{array}{c c c c}
e^{v}H_0& H_1         & 0     & 0\\
H_1     & e^{-v}{H_2} & 0     & 0\\
0       & 0           & r^2 K & 0\\
0       & 0           & 0     & r^2\sin^2\theta K
\end{array}
\r)Y_{lm} (\theta,\phi),
\label{eq:metrip}
\ee
where $(H_0,H_1,H_2,K)$ are functions of $(t,r)$ only. Additionally, the vector potential can be written as $A+\delta A$, where $A=-Q/r dt$ is the background vector potential of the RN BH. The perturbations $\delta A$ are decomposed as $\delta A=-(f_{02}dt + f_{12}dr)Y_{lm}$, with $f_{02}$ and $f_{12}$ being functions of $(t,r)$ only. The stress-energy tensor of the particle can also be decomposed in terms of spherical harmonics,
\be
T_{ab}^P=
\l(
\begin{array}{c c c c}
A_{(0)}                    & \frac{i}{\sqrt{2}} A_{(1)} &0 & 0\\
\frac{i}{\sqrt{2}} A_{(1)} & A                          & 0& 0\\
0                          & 0                          & 0& 0\\
0                          & 0                          & 0& 0
\end{array}
\r)Y_{lm} (\theta,\phi),
\label{eq:part}
\ee
where, once again, the $A$'s are functions of $(t,r)$ only. In a stationary background, we can explicitly eliminate the time dependence by a Fourier transformation of the perturbation functions:
we define the Fourier transform of a function $\psi(t,r)$ as 
\be
\Psi(\omega,t)=\frac{1}{\sqrt{2\pi}}\int dt e^{+i\omega t} \psi(t,r)\,.
\ee
Below we shall consider already the transformed Fourier quantities, using the same notation for each perturbation function.
By substituting the expressions for the metric and the vector potential into the Einstein equations and expanding up to first order, we obtain the following set of equations 
\bea
&&\frac{e^{2 v} {H_0}'}{r}-e^{2 v} {K}''-\frac{e^{2 v} {K}' \left(r v'+6\right)}{2 r}+\frac{{H_0} e^{v}}{2 r^4} \left[r^2 \left(l(l+1)+4 e^{v}-2\right)+2 Q^2+4 r^3 e^{v} v'\right]\nn\\
&+&\frac{\left(l(l+1)-2\right) {K} e^{v}}{2 r^2}=8 \pi  A_{(0)}-\frac{2 Q e^{v} {f_{02}}'}{r^2}-\frac{2 i Q \omega  {f_{12}} e^{v}}{r^2}\,,\label{eq:eintt}\\
&&i \omega  {K}'-\frac{i \omega  {H_0}}{r}-\frac{i \omega  {K} \left(r v'-2\right)}{2 r}+\frac{ {H_1} }{2 r^4}\left[r^2 \left(l(l+1)+2 e^{v}-2\right)+2 Q^2+2 r^3 e^{v} v'\right]\nn\\
&=&4 i \pi  \sqrt{2} A_{(1)}\,,\label{eq:eintr}\\	
&&\frac{1}{2} i \omega  {H_0}+\frac{1}{2} e^{v} {H_1}'+\frac{1}{2} {H_1} e^{v} v'+\frac{1}{2} i \omega  {K}=\frac{2 Q {f_{12}} e^{v}}{r^2}\,,\label{eq:eintth}\\	
&-&\frac{{H_0}'}{r}-\frac{2 i \omega {H_1} e^{-v}}{r}+{K}' \left(\frac{v'}{2}+\frac{1}{r}\right)+\frac{{H_0} e^{-v}}{2 r^4} \left[\left(l(l+1)-2\right) r^2+2 Q^2\right]\nn\\
&+&{K} \left(\omega ^2 e^{-2 v}-\frac{\left(l(l+1)-2\right) e^{-v}}{2 r^2}\right)=8 \pi  A+\frac{2 Q e^{-v} {f_{02}}'}{r^2}+\frac{2 i Q \omega  {f_{12}} e^{-v}}{r^2}\,,\label{eq:einrr}\\
&&\frac{{H_0}'}{2}+\frac{1}{2} {H_0} v'+\frac{i \omega}{2}  {H_1} e^{-v}-\frac{{K}'}{2}=\frac{2 Q {f_{02}} e^{-v}}{r^2}\,,\label{eq:einrth}\\
&&\frac{1}{2} r^2 e^{v} ({K}''-{H_0}'')-r e^{v} {H_0}' \left(r v'+1\right)+\frac{1}{2} r {H_0} e^{-v} \left[r \omega ^2-e^{2 v} \left(r v''+v' \left(r v'+2\right)\right)\right]\nn\\
&-&ir^2 \omega {H_1}'-\frac{i r \omega }{2 } {H_1} \left(r v'+2\right)+\frac{1}{2} r e^{v} {K}' \left(r v'+2\right)\nn\\
&&+{K} \left[-\frac{Q^2}{r^2}+\frac{1}{2} r^2 \omega ^2 e^{-v}+\frac{1}{2} r e^{v} \left(r v''+v' \left(r v'+2\right)\right)\right]=-2 Q {f_{02}}'-2 i Q \omega  {f_{12}}\,,\label{eq:einthth}
\eea
whereas the perturbed Maxwell equations yield 
\bea
&&-r^2 {f_{02}}''-2 r {f_{02}}'+l (l+1) {f_{02}} e^{-v}-i r^2 \omega  {f_{12}}'-2 i r \omega  {f_{12}}+Q {K}'=-4 \pi  r^2 {j_0} e^{-v},\label{eq:ma1}\\
&&-i r^2 \omega  {f_{02}}'+{f_{12}} \left(r^2 \omega ^2-l (l+1) e^{v}\right)+i Q \omega  {K}=4 \pi  r^2 {j_1} e^{v}, \label{eq:ma2}\\
&&-i \omega  {f_{02}} e^{-v}-e^{v} {f_{12}}'-{f_{12}} e^{v} v'=4 \pi  {j_2}.\label{eq:ma3}
\eea
In the above equations we already used the fact that $H_2=H_0$, required by the $(\theta,\phi)$ component of the perturbed Einstein equations. 

Due to the Bianchi identities, not all of the above equations are independent. 
Let us first look into the Maxwell equations~\eqref{eq:ma1}--\eqref{eq:ma3}. Equation~\eqref{eq:ma1} is automatically satisfied, and is a consequence of \eqref{eq:ma1}--\eqref{eq:ma2}
and the continuity equation for the currents. Indeed, it is easy to see that, by defining $f_{12}=e^v\psi_e$ and manipulating Eqs.~\eqref{eq:ma2} and \eqref{eq:ma3}, we obtain Eq.~\eqref{eq:psie}.

Simplification of the gravitational sector is more involved~\cite{Zerilli:1974ai}. By manipulating Eqs.~\eqref{eq:eintr}--\eqref{eq:einrth} we can obtain a system of two differential equations for $H_1$ and $K$, and an algebraic relation between $H_0$, $H_1$, and $K$. By defining the vector $\bm \psi=(K,-i\omega^{-1}H_1)$, the differential equations can be written as
\be
\l(\frac{d}{dr} -\bm A \r)\bm\psi=\bm S\,,	
\ee
where $\bm A=
\l(
\begin{array}{c c}
\alpha &\beta\\
\gamma &\delta
\end{array}
\r)$, $\bm S=(S_1,S_2)$, and the coefficients $(\alpha,\beta,\gamma,\delta)$ are linear in $\omega^2$, e.g. $\alpha=\alpha_0+\omega^2\alpha_2$. To simplify the system further, we wish to find a transformation
\be
\bm \psi = \bm F \bar{\bm \psi}+\bar{\bm S}\,,
\ee
with $\bm F=
\l(
\begin{array}{c c}
f & g\\
h  & k 
\end{array}
\r)$ and $\bar{\bm \psi}=(\psi_g,\psi_g')$, such that the function $\psi_g$ obeys the following differential equation 
\be
n(n\psi_g' )' +(\omega^2-V_g)\psi_g=S_{z}.
\ee
Note that $S_z$ will also involve EM perturbations, since these appear as a source terms in the first-order equations. By comparing the coefficients of different order in $\omega$ in the differential equations it is possible to obain a relation involving $(f,g,h,k,n,V_g,S_z)$ and their derivatives. We obtain $g=1$, $n=e^v$, $k=-r e^{-v}$, and
\bea
f&=&\l[r \left(M+r \left(-\Lambda +2 e^{v}-2\right)\right)\r]^{-1}\bigg\{(\Lambda +1) [M-(\Lambda +2) r]+e^{v} [2 (\Lambda +3) r-5 M]-4 r e^{2 v}\bigg\},\nn\\
h&=&\frac{3 M+2 \Lambda  r}{M+r \left(-\Lambda +2 e^{v}-2\right)}+\frac{(r-M) e^{-v}}{r}\,.
\eea
The remaining functions are given below.

\subsection{Final equations}
Through the procedure described in the previous section, we obtain a system of two coupled second-order differential equations, namely Eqs.~\eqref{eq:psig} and \eqref{eq:psie}, where the coefficients are given by
%
\begin{align}
V_g&=\frac{2 \left(r (r-2 M)+Q^2\right) }{r^6 \left(r (3 M+\Lambda  r)-2 Q^2\right)^2}\bigg[Q^2 r^2 \left(-27 M^2+2 (3-8 \Lambda ) M r+6 \Lambda  r^2\right)\nn\\
&+r^3 \left(9 M^3+9 \Lambda  M^2 r+3 \Lambda ^2 M r^2+\Lambda ^2 (\Lambda +1) r^3\right)+6 Q^4 r (3 M+\Lambda  r)-4 Q^6\bigg],\\
V_e&=\frac{2 e^{v} \left(Q^2 \left(\Lambda -4 e^{v}+1\right)+(\Lambda +1) r^2 \left(-2 \Lambda +3 e^{v}-3\right)\right)}{r^2 \left(Q^2+r^2 \left(-2 \Lambda +3 e^{v}-3\right)\right)},\\
	I_1&=\frac{8 i Q e^{v} \left(Q^4-2 Q^2 r^2 \left(2 e^{v}+1\right)+r^4 \left(-4 \Lambda ^2-8 \Lambda +3 e^{2 v}-3\right)\right)}{r^3 \omega  \left(Q^2+r^2 \left(-2 \Lambda +3 e^{v}-3\right)\right)^2},~I_3=\frac{i Q \omega  e^{2 v} }{r^2},\\
I_2&=\frac{i Q \omega  e^{v} \left(Q^2 \left(-\Lambda +5 e^{v}-1\right)+r^2 \left((\Lambda +1) (2 \Lambda +3)-3 (\Lambda +2) e^{v}+3 e^{2 v}\right)\right)}{r^3 \left(Q^2+r^2 \left(-2 \Lambda +3 e^{v}-3\right)\right)},\\
S_g&=-\frac{16 i \pi  Q e^{2 v} r j_1}{\omega  \left(Q^2+r^2 \left(-2 \Lambda +3 e^{v}-3\right)\right)}-\frac{16 \pi  r^3 A e^{2 v}}{Q^2+r^2 \left(-2 \Lambda +3 e^{v}-3\right)}
-\frac{8 \sqrt{2} \pi  r^3 e^{2 v} A_1'}{\omega  \left(Q^2+r^2 \left(-2 \Lambda +3 e^{v}-3\right)\right)}\nn\\
&-\frac{8 \sqrt{2} \pi  A_{1} e^{v} \left(-Q^4+2 Q^2 r^2 \left(\Lambda +3 e^{v}+2\right)+r^4 \left(e^{v}-1\right) \left(2 \Lambda +3 e^{v}+3\right)\right)}{\omega  \left(Q^2+r^2 \left(-2 \Lambda +3 e^{v}-3\right)\right)^2},\\
S_e&=4 \pi j_1 e^{2 v}-\frac{8 i \sqrt{2} \pi  Q r A_{1} e^{2 v}}{Q^2+r^2 \left(-2 \Lambda +3 e^{v}-3\right)},
\end{align}
%
with
\begin{align}
	\Lambda&=\frac{1}{2}(l-1)(l+2),\\
	j_1&=-\frac{\sqrt{1+2l} \,q e^{i \omega  T}}{2 \sqrt{2} \pi  (r-r_-) (r-r_+)},\\
	A_1&=\frac{i (q Q-r\gamma)  \sqrt{1+2l}\,\mu e^{i \omega  T}}{2 \pi r  (r-r_-) (r-r_+)},\\
	A&=-\frac{(q Q-r\gamma)  \sqrt{1+2l}\,\mu r^3 e^{i \omega  T}}{2 \sqrt{2} \pi  (r-r_-)^3 (r-r_+)^3 T'},
\end{align}
where $T(r)$ is the function that describes the particle motion, $q$ is the particle charge, $\gamma$ is the energy per unit of mass as measured by an asymptotic observer, and $r_+$ ($r_-$) is the outer (inner) horizon of the Reissner-Nordstrom BH, $r_\pm=M\pm\sqrt{M^2-Q^2}$. Note that a point charge will not follow a geodesic of the background metric and its motion is defined by~\cite{Chandrasekhar:1985kt}
\begin{align}
T'&=\frac{q Q-\gamma  r}{e^v \sqrt{(q Q-\gamma  r)^2-r^2e^v}},\\
\frac{dt}{d\tau}&=\frac{r\gamma-qQ}{r e^v}\,.
\end{align}
The equation for $T(r)$ can be integrated analytically, but the result is cumbersome and we don't show it.
\subsection{Numerical procedure}
To solve the perturbation equations~\eqref{eq:psig} and \eqref{eq:psie} we employed two different methods. The latter agree with each other within numerical accuracy. 

The first method relies on the Green's function approach, also called method of variation of parameters~\cite{boyce2008elementary}. Let $\bm \Psi =(\psi_g,\psi_e,d\psi_g/dr_*,d\psi_e/dr_*)$. The system~\eqref{eq:psig}--\eqref{eq:psie}  can be written as
\be
\frac{d}{dr_*}\bm\Psi+\bm V\bm \Psi=\bm S.
\ee
We start by constructing the fundamental $4\times 4$ matrix $\bm X$, whose columns are independent solutions of the associated homogeneous differential equations. The independent solutions can be obtained in the following way. We notice that at the horizon, the required solutions have the following form:
\be
\psi_{g,e}\sim A^{r_+}_{g,e} e^{-i\omega r_*}\,,
\label{eq:bounh}
\ee
and at infinity we have
\be
\psi_{g,e}\sim A^{\infty}_{g,e} e^{i \omega r_*}\,.
\label{eq:bouni}
\ee
In the above equations $A^{r_+,\infty}_{g,e}$ are constants.
The first two columns of $\bm X$ can be obtained by integrating the homogeneous equations from the horizon outwards with boundary conditions \eqref{eq:bounh}, by choosing two independent solutions, say $(A^{r_+}_g,A^{r_+}_e)=(1,0)$ and $(A^{r_+}_g,A^{r_+}_e)=(0,1)$. The other two columns of the matrix $\bm X$ can be obtained by integrating from  infinity inwards, with boundary conditions~\eqref{eq:bouni}, by choosing, once again, $(A^{\infty}_g,A^{\infty}_e)=(1,0)$ and $(A^{\infty}_g,A^{\infty}_e)=(0,1)$. The general solution can be written in terms of the fundamental matrix as
\be
\bm \Psi=\bm X\int dr_* \bm X^{-1} \bm S := \bm X {\bm I	}\,.
\ee
The limits of the above integral are suitably chosen such that the general solution obeys the proper boundary conditions at the horizon and at infinity, i.e., ingoing waves at the horizon and outgoing waves at infinity. From the final solution, the asymptotic form of the gravitational and EM master functions is given by
\begin{eqnarray}
 \psi_g &\sim& e^{i \omega r_*}I_3\,,\\
 \psi_e &\sim& e^{i \omega r_*}I_4\,,
	\end{eqnarray}
and therefore the GW and EM fluxes can be obtained through the absolute value of the integral component $I_3$ and $I_4$, respectively [cf. Eqs.~\eqref{GWflux}--\eqref{EMflux}]. At the horizon, $\psi_g\sim e^{-i \omega r_*} I_1$ and $\psi_e\sim e^{-i \omega r_*} I_2$. 

Let us now expose the second method that we used to solve the system~\eqref{eq:psig}--\eqref{eq:psie}. The shooting method relies on an integration of the full system of inhomogeneous equations. First, we impose that the solutions near the horizon are of the form \eqref{eq:bounh}. We then integrate the full equations up to infinity where the general solution will be a superposition of the ingoing and outgoing modes, namely
\be
\psi_{g,e}\sim A_{g,e}^{in}e^{-i \omega r_*}+A_{g,e}^{out}e^{i \omega r_*}.
\ee
The physical solutions corresponding to a particle falling into the BH require $A_{g,e}^{in}=0$, and, therefore, this becomes a two-parameter shooting problem for the amplitudes $A_{g,e}^{r_+}$. With the proper values of $A_{g,e}^{r_+}$, we can compute the amplitude of the GW and EM waves at infinity, which enable us to compute the GW and EM energy spectra through Eqs.~\eqref{GWflux}--\eqref{EMflux}.

\subsection{Supplemental results}
\begin{figure*}[ht]%
\begin{center}
\includegraphics[width=0.5 \textwidth]{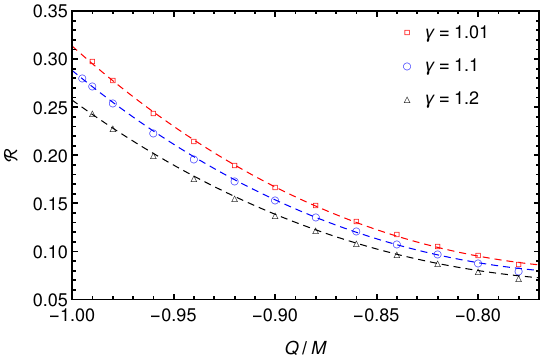}%
\caption{Ratio between the two peaks of the quadrupolar GW flux due to the excitation of both the EM and the gravitational mode of a Reissner-Nordstrom BH during the plunge of a charged particle with opposite charge. The dashed lines show a cubic fit in the variable $Q/M$, cf. Eq.~\eqref{ratio}.}%
\label{fig:ratio_f}%
\end{center}
\end{figure*}

In this section we present some supplemental results on the GW and EM emission in the radial infall of a point charge into a Reissner-Nordstrom BH. 
Figure~\ref{fig:ratio_f} shows the ratio between the two peaks of the quadrupolar GW flux [cf. Eq.~\eqref{ratio}] as a function of the BH charge. As discussed in the main text, the relative amplitude of the EM peak is nonnegligible only when $Qq<0$ and when the BH is highly charged, cf. Fig.~\ref{fig:spectrum}. Note that the ratio depends on the boost factor $\gamma$.

\begin{figure*}[ht]
\begin{center}
\epsfig{file=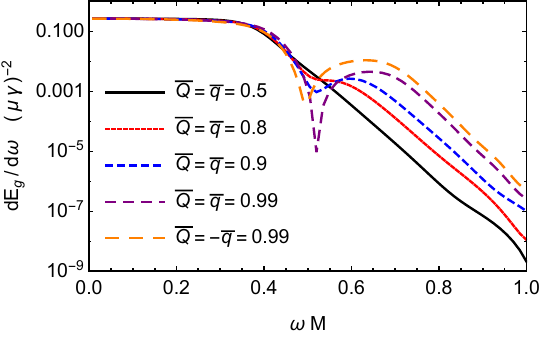,width=7.8cm,angle=0,clip=true}\epsfig{file=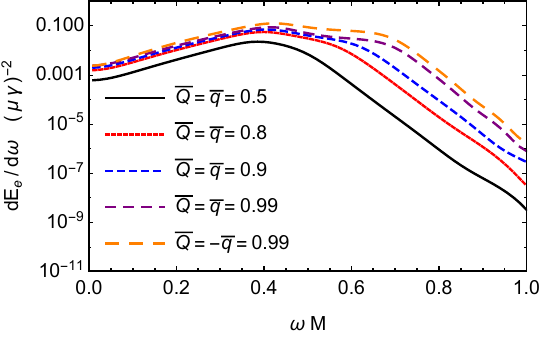,width=7.8cm,angle=0,clip=true}
\caption{
Quadrupolar GW (left panel) and EM (right panel) energy spectra for a high-energy charged particle plunging radially on a RN BH. The particle and the BH have the same charge-to-mass ratio. 
In the legend of both panels $\bar{Q}:=Q/M$ and $\bar{q}:=q/\mu$.
\label{fig:spectrum_high}}
\end{center}
\end{figure*}

Finally, for completeness in Fig.~\ref{fig:spectrum_high} we present some representative cases for the GW and EM energy spectra for the radial infall of a high-energy point charge, i.e. $\gamma\to\infty$. In this regime our results for the EM flux are in perfect agreement with those presented in Ref.~\cite{Cardoso:2003cn}.

\bibliographystyle{JHEP}
\bibliography{references}

\end{document}